\documentclass[preprint]{elsarticle}
\usepackage{graphicx}
\usepackage{dcolumn}
\usepackage{bm}
\usepackage{mathrsfs}
\usepackage{amsmath}
\usepackage{amssymb}
\usepackage{geometry}
\usepackage{xr}
\usepackage{xcolor}
\usepackage{soul}
\usepackage{url}

\newif\ifshowedit
\newif\ifshoweditTwo
\newif\ifshoweditThree

\showeditfalse 

\showeditTwofalse 

\showeditThreefalse 

\ifshowedit
	 \newcommand{\add}[1]{{\color{blue}#1 }}
 	\newcommand{\del}[1]{{\color{red} \st{#1}}}

 	\newcommand{\deleq}[1]{{\color{red} \setstcolor{red}#1}}
\else
	\newcommand{\add}[1]{#1}
	\newcommand{\del}[1]{}
	\newcommand{\deleq}[1]{}
	
 \fi
\ifshoweditTwo
	 \newcommand{\addTwo}[1]{{\color{blue}#1 }}
	\newcommand{\delTwo}[1]{{\color{red} \st{#1}}}
 	\newcommand{\deleqTwo}[1]{{\color{red} \setstcolor{red}#1}}
\else
 	\newcommand{\deleqTwo}[1]{}
	\newcommand{\addTwo}[1]{#1}
	\newcommand{\delTwo}[1]{}
 \fi

\ifshoweditThree
	 \newcommand{\addThree}[1]{{\color{blue}#1 }}
	\newcommand{\delThree}[1]{{\color{red} \st{#1}}}
 	\newcommand{\deleqThree}[1]{{\color{red} \setstcolor{red}#1}}
\else
 	\newcommand{\deleqThree}[1]{}
	\newcommand{\addThree}[1]{#1}
	\newcommand{\delThree}[1]{}
 \fi

 \title{The Bispectrum and Its Relationship to Phase-Amplitude Coupling }

\author[uihc]{Christopher K. ~Kovach\corref{cor1}}
\author[uihc]{Hiroyuki ~Oya}
\author[uihc]{Hiroto ~Kawasaki}

\cortext[cor1]{Corresponding author}

\address[uihc]{
 Department of Neurosurgery\\
 University of Iowa Carver College of Medicine\\
 Iowa City, Iowa
 }

\date{\today}

\begin{document}

\begin{abstract}

\add{
Most biological signals are non-Gaussian, reflecting their origins in highly nonlinear physiological systems. 
A versatile set of techniques for studying non-Gaussian signals relies on the spectral representations of higher moments, known as polyspectra, which describe forms of cross-frequency dependence that do not arise in time-invariant Gaussian signals.}\add{
The most commonly \addThree{used of these} employ the bispectrum. 
Recently, other measures of cross-frequency dependence have drawn interest in EEG literature, in particular those which address phase-amplitude coupling (PAC).}\add{
Here we demonstrate a close relationship between the bispectrum and popular measures of PAC, which we relate to}
smoothings of the signal bispectrum, making them fundamentally bispectral estimators.
Viewed this way, however, conventional PAC measures exhibit some unfavorable qualities, including poor bias properties, lack of correct symmetry and artificial constraints on the spectral range and resolution of the estimate.
Moreover, information obscured by smoothing in measures of PAC, but preserved in standard bispectral estimators, may be critical for distinguishing nested oscillations from  transient signal features and other non-oscillatory causes of ``spurious'' PAC.
\add{
We propose guidelines for gauging the nature and origin of cross-frequency coupling with bispectral statistics.}
Beyond clarifying the relationship between PAC and the bispectrum, \add{the present work lays out a general framework for the interpretation of the bispectrum, which extends to other higher-order spectra.
In particular, this framework holds promise for the detailed identification of signal features related to both nested oscillations and transient phenomena.
We conclude with a discussion of some broader theoretical implications of this framework and highlight promising directions for future development.} 

\end{abstract}


\maketitle

\section{Introduction}

Many interesting properties of signals in nature relate to nonlinear, non-Gaussian and non-stationary dynamics, which are poorly indexed by second-order measures such as power and cross spectra.
Because the spectrum of a stationary Gaussian process lacks any statistical dependence across frequencies, measures of frequency-domain dependence can be extremely useful for gauging the presence and nature of higher order dynamics \cite{Sheremet4218}.
For stationary signals, higher order spectra, ``polyspectra,'' which capture such  dependence, are the frequency-domain representations of higher moments \citep{nikias1993signal}, in which they mirror the relationship between the power spectrum of a signal and its autocorrelation.
The bispectrum is the third-order polyspectrum, making it an obvious place to start in the approach towards higher-order dynamics, at least from a statistical standpoint.
Bispectral analysis has proved its practical worth in applications to EEG \citep{dumermuth1971analysis,sigl1994introduction}, most notably in gauging the depth of anesthesia \citep{barnett1971bispectrum,gan1997bispectral,kearse1998bispectral,myles2004bispectral}. 
Nevertheless, a significant drawback for the non-statistician  remains its lack of an obvious connection to any simple physical interpretation \citep{fackrell1995interpretation}. 
This point has been the source of some confusion in applied literature; for example, as recently reviewed by \citet{hyafil2015misidentifications}, various authors have suggested incorrectly that bicoherence relates to phase entrainment across pairs of frequencies. 
While it is true that bispectral measures cannot easily be reduced to any single interpretation, a goal of the present work is to show that specific forms of dependence leave easily recognized signatures in the bispectrum, making bispectral measures invaluable for distinguishing between a variety of phenomena related to cross-frequency coupling.

A separate body of work has recently emerged from EEG literature, which examines the role of another form of cross-frequency coupling, phase-amplitude coupling (PAC) \citep{canolty2006high,hyafil2015neural}. PAC refers to dependence between analytic amplitude at one frequency and analytic phase at another.
Rather than any statistical first principle, interest in PAC has been motivated by the empirical discovery of PAC in signals recorded from the brain, alongside emerging computational and physiological models of how it arises within populations of interacting neurons \citep{young2009coupling,jirsa2013,akam2014oscillatory,hyafil2015neural}. 
In this literature, the measurement of PAC is usually approached with second-order statistics, such as coherence, applied towards comparing analytic phase in one band with analytic amplitude extracted from another, treating the two as separate signals  \citep{schack2002phase,canolty2006high, Tort2010}. 
The situation with PAC is therefore reversed from bispectral measures: its physical meaning is more evident than its relationship to any general body of statistical theory.

The present work aims to bridge this gap and resolve ambiguities of meaning in both directions.  
The central result is that second-order measures of PAC may be fundamentally understood as estimators of the bispectrum.
In the same way that windowed stationary\del{ and time-varying} estimates of the power spectrum can be equated to smoothings of the \del{Wigner-Ville distribution}\add{true signal spectrum}, windowed bispectral estimators, which include those underlying measures of PAC, amount to different ways of smoothing \del{and integrating over the third-order Wigner-Ville distribution}\add{the true signal }\addTwo{bispectrum}\delTwo{polyspectra} \citep{gerr88,nikias1987bispectrum,swami88,hanssen2003theory}. In both cases, differences between estimators relate to properties of the respective smoothing kernels \citep{Cohen1989}. 
This observation demonstrates conclusively the meaning of PAC measures as they relate to the bispectrum and vice versa and establishes that second-order measures of PAC provide no unique information beyond what can be obtained from the bispectrum.

While PAC measures are fundamentally measures of the bispectrum, the reverse is not true; it is not correct to conclude that the bispectrum is principally a reflection of phase-amplitude coupling. 
Forms of ``spurious'' PAC,  related, for example, to spectrally broad signal features, may be traced to the bispectral nature of PAC measures. 
One practical implication is that the superior resolution and lower bias of standard bispectral measures, in comparison to PAC measures,  allows them to retain information critical for distinguishing between nested oscillations and other sources of apparent phase-amplitude coupling \citep{kramer2008sharp}.
Following a brief review of bispectral and PAC estimation, we will observe how different regions of the bispectrum may be taken to reflect either phase-amplitude coupling or consistency of phase across a range of frequencies. 
It is shown that the bispectrum can be highly useful for ascertaining the presence and origin of phase-amplitude coupling.
Properly applied and interpreted, bispectral statistics may overcome a number of recently highlighted limitations and ambiguities of existing PAC measures\cite{aru2015untangling,hyafil2015misidentifications,kramer2008sharp,lozano2016neuronal,Sheffer-Teixera16,van2015phase}.
For example, with conventional measures of PAC, the observable range of phase-providing frequencies is restricted by the bandwidth of the amplitude-providing band \cite{aru2015untangling}, yet no such limitation applies to bispectral estimates. 
In light of the relationship to the bispectrum, it becomes clear that this constraint is an artifact of the estimator rather than anything inherent in the quantity measured.

\add{Many of the questions that arise in considering the relationship between PAC and the bispectrum prove to be of much more general relevance for understanding a range of signal properties that are neglected by traditional spectral measures, a topic whose importance is becoming increasingly clear \citep{cole2017brain}.}\add{
In particular, we develop a model that uses the bispectrum to capture spectrally complex signal features, of which PAC is only one example.}\add{
The concluding sections are devoted to a prospective review of the application of the bispectrum towards understanding the nature and functional significance of non-oscillatory and transient sources of cross-frequency coupling, beyond nested oscillations.}

\subsection{\add{Organization}}
\label{section:organization}
\add{
The overall aim of the current work is threefold; first, a general introductory background is provided to motivate applications of higher-order spectra in signal analysis, accompanied in the appendices by a more focused and technical review of the bispectrum and its estimation.}\add{
The second aim is to describe a formal equivalence between bispectral estimators and measures of phase-amplitude coupling, details of which are presented in \ref{appendix:PhPC_proof}.}\add{
Finally, building on this formal relationship\delTwo{s}, a framework is developed to guide the interpretation of the bispectrum.}\add{
In most places, more technical development has been left to the appendices, the results of which are summarized in the main text alongside some background explanation.

Section \ref{SEC_Primer}, in particular, focuses on introducing readers who have had little exposure to the ideas and applications of higher-order spectra to some general motivating principles, thus it deals with the subject very broadly.
Section \ref{SEC_bispectrum} gives a short summary of the main relevant properties of the bispectrum described in more detail in \ref{section:bispectral_properties} and \ref{section:bispectral_estimators}.}\add{
Section \ref{SEC_PAC} similarly introduces phase-amplitude coupling, which is given a more detailed treatment in \ref{appendix:PhPC_proof}. 
Section \ref{section:bispectral_signatures} develops a framework for understanding the relationship between PAC and the bispectrum, building on the proof provided in \ref{appendix:PhPC_proof}.}\add{ 
Section \ref{section:FM_bispect} briefly considers an extension to phase-frequency coupling. 
The concluding sections set the framework within a broader theoretical context; section \ref{section:encoding} describes a model of signal encoding and \delThree{transduction}\addThree{transmission} that exploits properties of the bispectrum, while section \ref{sec:future} anticipates further development of the framework, especially in the direction of feature extraction and identification.}

\section{\add{A primer on higher-order spectra in signal processing}}
\label{SEC_Primer}
\add{
Abstractly, signal processing is about making sense of sequential or otherwise meaningfully ordered quantities; it considers the question, what kinds of measures best capture the essential nature of any dependence among the quantities as it relates  to their position in the ordering space (henceforth assumed to be time)?}
\add{An endless variety of different measures can be conceived, the usefulness and interpretation of any of which naturally depends on very concrete particulars of the application.}\add{
But to a theoretically-minded statistician, armed with the hammer of distribution theory, a signal is a nail drawn from a distributional pile, and signal processing is merely the application of a general form of statistical inference to the special case of ordered data.}\add{
Our slightly naive theoretician might be forgiven for turning to measures related to \emph{moments}, because, at least in princip\delTwo{al}\addTwo{le}, moments completely characterize a distribution, meaning that they collectively harbor all the information there is to harbor about the process generating the signal, irrespective of its source or nature.}

\add{
Moments are most familiar in the form of the mean, variance, skewness and kurtosis of univariate distributions, related to the expected values of the first four powers of the random variable.  
The idea generalizes to the expected value of any power; the $k^{\mathrm{th}}$-order moment is the expectation of the $k^{\mathrm{th}}$ power. 
The extension to multivariate distributions adds cross-terms to the picture, which are most commonly encountered \delTwo{in the form of}\addTwo{in} the off-diagonal values of a covariance matrix.
At higher orders, cross-terms within the $k^{\mathrm{th}}$ moment may take the form of the product of $k$ distinct variates or any combination of powers of the variates that sum to $k$.
For example, the $k^{\mathrm{th}}$ moment of a distribution over $(x,y,z)$ includes all of the combinations of powers that appear in the expansion of $(x+y+z)^k$.
}\add{

In applications to signal processing, the sampling space is neither univariate nor in general multivariate, rather it is the space of possible signals, which, for theoretical purposes, occupies a continuously infinite number of dimensions.
This fact causes moments to take on a decidedly intimidating air in signal-processing contexts, but the truth is that they behave more-or-less exactly like their tamely finite cousins. 
The ordering space of the signal (in our case, time) takes the place of the integer-valued index of multivariate distributions. 
In analogy to the cross-terms of multivariate distribution, moments of a continuous signal involve the expected values of the products of the signal with itself at specified points in time.}\add{
For example, if $X(t)$ is a bounded signal, meaning here the relevant moments are defined, its first moment at time $t_1$ is the expectation of the signal itself, $\mathrm{E}\left[X(t_1)\right]$,  the second moment at $(t_1,t_2)$ is $\mathrm{E}\left[X(t_1)X(t_2)\right]$, while the expectation of the product 
\begin{align}
\label{EQ_moment}
\add{\mathrm{E}\left[X(t_1)X(t_2)...X(t_k)\right]} 
\end{align}
is the $k^\mathrm{th}$ moment at $\left(t_1,..,t_k\right)$.} 

\add{
In spite of their considerable theoretical appeal, the application of moment-related statistics to real-world problems in signal processing invariably runs into two obstacles.
First, their estimation puts severe demands on data because moments may assume a set of distinct values over a parameter space that grows exponentially with order, starting from one that is already at least as large as the observed signal.
Second, the question of how to interpret higher moments also seems to become exponentially less clear with increasing order. 
It is therefore impossible to estimate moment-related statistics without some judicious assumptions that rein in the parameter space, but the exercise is still bound to prove hollow if one lacks any insight into real-world meaning.}

\add{
The first point is addressed easily enough by making assumptions about how the dependence relates to ordering; for example, windowing is commonly used to enforce the assumption that any dependence grows negligible outside some local neighborhood according to the window.
Combined with the assumption of time invariance (reviewed in the following section) windowing and similar procedures make estimation of higher moments a tractable problem.   
Even so,\add{ the }practical application \addTwo{of moment-related statistics in signal-processing contexts} rarely involves orders above 4, while applications involving 3rd order dependence, of the type that will be given special attention here, already border on exotic.
Moment-related statistics of order 1 and 2 are perfectly familiar as mean and covariance or autocorrelation;
the leap from 2 to 3 introduces qualitatively new properties with potentially far-reaching implications, a few of which we will encounter in the following discussion. }

\subsection{\add{Time invariance}}
\label{section:time_invariance}
\add{A common and sometimes necessary simplifying assumption is that the statistics governing the signal remain constant over time. 
A very useful byproduct of time invariance, and motivation for giving special consideration to invariant measures, is that estimation may resort to averaging over time; for example, the estimate of the fixed first moment becomes the average $\hat{\mu}_1=\frac{1}{T}\int_0^T{X(t)\mathop{dt}}$, which ought to converge with the true expectation, $\mu_1$, as $T$ increases (provided again the signal is suitably bounded). 
Similarly, time invariance allows the second moment to vary only with respect to the delay between two points in time, $\mu_2(\tau)=E\left[X(t)X(t+\tau)\right]$. The second moment may therefore be estimated by averaging  over time in the same way:
\begin{align}
\label{EQ_autocorr}
\hat{\mu}_2(\tau)=\frac{1}{T}\int_0^T{X(t)X(t+\tau)\mathop{dt}}
\end{align}
giving the familiar autocorrelation. 
Likewise, for higher moments, time-invariant statistics are those obtained by considering the lags between the indexed times, collapsing over the absolute time. }

\subsection{\add{Spectra and time-invariant moments}}
\add{A central result in signal-processing theory concerns the relationship between time-invariant statistics and the spectral representation of the signal, given by its Fourier transform: $\tilde{X}(\omega) = \mathscr{F}\left\{X(t)\right\}$.} \add{
From the inverse Fourier transform:}
\add{
\begin{align}
\label{EQ_spectInv}
	X(t) &= \mathscr{F}^{-1}\left\{\tilde{X}\left(\omega\right)\right\} = \frac{1}{2\pi}\int{\tilde{X}(\omega)e^{i\omega t}\mathop{d\omega}}
\end{align}
one can readily appreciate that any shift of time, $\Delta t$, separates out into an exponential term in the spectrum of $X$:  
\begin{align}
\label{EQ_spectral_tshift}
\mathscr{F}\left\{X(t+\Delta t)\right\} = \tilde{X}(\omega)e^{i\omega\Delta t}
\end{align}}\add{
Consider the effect of a random shift\addTwo{, $\Delta t$}, with uniform distribution (and otherwise independent of the unshifted signal) on the second moment:}
\add{
\begin{align}
\begin{split}
\label{EQ_spectexp}
	\mu_\addTwo{2}(\tau)&=\mathrm{E}\left[X(t+\Delta t)X(t+\Delta t + \tau)\right] \\
		&= \frac{1}{2\pi}\int{\mathrm{E}\left[\tilde{X}(\omega_1)\tilde{X}(\omega_2)\right]e^{i(\omega_1+\omega_2)t+i\omega_2 \tau}\mathrm{E}\left[e^{i(\omega_1+\omega_2) \Delta t}\right]\mathop{d\omega_1d\omega_2}}\\
	&=\frac{1}{2\pi}\int{\mathrm{E}\left[\tilde{X}(\omega_1)\tilde{X}(\omega_2)\right]e^{i(\omega_1+\omega_2)t+i\omega_2 \tau}\delta(\omega_1+\omega_2)\mathop{d\omega_1d\omega_2}}\\
	&=\frac{1}{2\pi}\int{\mathrm{E}\left[\left|\tilde{X}(\omega_2)\right|^2\right]e^{i\omega_2\tau}\mathop{d\omega_2}}
\end{split}
\end{align}
In other words, the time-invariant second moment, $\mu_2(\tau)$, is the time-domain representation of the power spectrum, a result known as the Wiener-Khintchine theorem (technically speaking, what is given here is a special case valid for integrable functions).}
 
\add{More generally, paralleling the situation in the time domain, moments are represented in the spectral domain as the expectation of products of the signal spectrum across different frequencies. 
One can conclude from a glance at Eq. (\ref{EQ_spectral_tshift}) that any time-invariant products must arise from those combinations of frequencies for which the exponential terms cancel. 
For the first moment, the exponential term vanishes only at $\omega=0$, hence time invariance applies only at the origin, $\tilde{X}(0)$, related to the fixed constant mean.
For the second moment, the exponent vanishes when $\omega_1=-\omega_2$, as then $\tilde{\mu}_2=\tilde{X}(\omega)e^{i\omega\Delta t}\tilde{X}(-\omega)e^{-i\omega\Delta t}=\left|\tilde{X}(\omega)\right|^2$. 
The idea extends naturally to higher moments as well; in general, given the product
\begin{align}
\begin{split}
\label{EQ_HOS}
	\tilde{\mu}_{K} &= \mathrm{E}\left[\tilde{X}(\omega_1)\tilde{X}(\omega_2)\ldots\tilde{X}(\omega_K)\right]\\
\end{split}
\end{align}
time invariant statistics relate to those terms for which
	\[\sum_{i=1}^{K}\omega_i=0\] }\add{
Such products go by the names \emph{higher order spectra} or \emph{polyspectra} and it is not very difficult to show that, as with the second-order case, they are related to the time-domain moment of the corresponding order by a multi-dimensional Fourier transform. 
The 3rd order polyspectrum is known as the \emph{bispectrum} while the 4th order is the \emph{trispectrum}.}

\subsection{\add{Time-Varying Spectra and the Wigner-Ville Distribution}}

\add{In the previous section we equated time-invariant spectra to the spectral representation of a time-varying moment integrated over time, which we may likewise interpret as an estimate of the time-invariant spectrum obtained by averaging over time. 
But there are plenty of applications in which it must be assumed that\delTwo{the power spectrum varies}\addTwo{spectra vary} over time. 
If the time-invariant estimate averages over time, shouldn't it be possible to observe the time-varying spectrum simply by not averaging?
An important class of time-frequency representations does just that, retaining the Fourier transform but skipping the integration over time; under a suitable change of variables, the outcome represents the original time-varying moment as a time-frequency distribution.
The best-known example of such an object is the Wigner-Ville distribution (WVD):}
\add{
\begin{align}
\begin{split}
\label{EQ_WigVille}
	W(\omega,t) &= \int{X\left(t+\frac{\tau}{2}\right)X\left(t-\frac{\tau}{2}\right)e^{-i\omega\tau}\mathop{d\tau}}
\end{split}
\end{align}}\add{ 
The WVD applies a Fourier transform along the lag dimension of the time-varying second moment, combined with a trivial change of variables that centers the lag, $\tau$, on absolute time, $t$.
Though sometimes used as a time-frequency representation in its own right, it suffers from the drawback that ``spurious'' energy may arise as a result of interference between spectrally isolated components in regions where neither component separately contains energy. 

The WVD is, arguably, more useful as a conceptual starting point in developing a general theory of power-spectral estimation: any of the common stationary and time-varying Fourier-derived power spectra can be gotten by smoothing or integrating over the distribution, which has the effect of smoothing away energy related to interference \citep{hlawatsch1992linear,Cohen1989}.
Distinctive properties of any estimator are a consequence of the particular window it applies to the distribution.
This idea generalizes naturally to higher moments, whose estimators may be likewise equated to smoothings of higher-order Wigner-Ville distributions, with distinguishing properties set by the choice of smoothing window.  }

\subsection{\add{Polyspectra and Gaussian Processes}}
\add{One of the important qualitatively new properties of third-order and higher polyspectra arises from the fact that Gaussian signals are fully characterized by second-order statistics, meaning that any linear time-invariant (LTI) Gaussian process is completely determined by its mean and power spectrum.} 
\add{All forms of cross-frequency dependence relate to polyspectra of order 3 or higher (those with two or more frequency dimensions), therefore LTI Gaussian processes  exhibit no dependence across frequencies. 
One of the main selling-points of higher-order spectra is that they can reveal the presence of non-Gaussian behavior as might result, for example, from non-linear signal dynamics. 
At the conclusion of this work we will see that it is possible to isolate such non-Gaussian signals from a background of Gaussian noise on the basis of higher-order spectra.
This is an especially compelling application of higher-order spectra because the most interesting signal components are also often the least Gaussian. 
} 
\subsubsection{\addTwo{Moments and Cumulants}}
\addTwo{Cumulants are closely related to moments, but for Gaussian processes they are identically zero at all orders above two.
They may therefore be constructed by subtracting the moment of a Gaussian process with a matching spectrum from the corresponding signal moment. 
For higher-order spectra, this generally involves subtracting a term within those sub-domains wherein the polyspectrum reduces to a product of power spectra and/or any non-zero mean. 
For example, the \delThree{the} cumulant \emph{tri}spectrum subtracts $\mathrm{E}\left[\left|\tilde{X}(\omega_1)\right|^2\right]\mathrm{E}\left[\left|\tilde{X}(\omega_2)\right|^2\right]$ from the moment trispectrum within the subspace where $\omega_3=-\omega_1$, while
outside this region the moment and cumulant trispectra are equal. 
The cumulant bispectrum and moment bispectrum diverge only along $\omega_1=0$, $\omega_2=0$ and $\omega_1=-\omega_2$ for a non-zero mean process but are identical everywhere for a zero-mean process, a property that extends to all odd-ordered cumulants.}

\subsection{\add{Cross Polyspectra}}
\add{In the polyspectral moment products described by Eq (\ref{EQ_HOS}), each term is drawn from the same signal, but there is nothing to stop us from mixing terms across multiple signals. 
Such cross polyspectra, which are the higher-order generalizations of the cross spectrum, describe various forms of dependence between multiple signals and are cross terms in the moments of the corresponding multivariate signal distributions. 
The practical application of these measures to EEG has gained some recent attention \citep{shils1996bispectral,chella2014third}. 

Third-order cross polyspectra also arise implicitly when considering the dependence between second-order statistics, and independent measures of interest.
For example, it is often of interest to relate ordinary power spectra or cross spectra to experimental variables that change over time. 
Standard regression models which treat power as the dependent measure implicitly model a subset of cross-bispectral interactions with the independent variable.
The possible value of a more explicit treatment of third-order interactions in such analyses is a question worth pondering.
} 

\subsection{\add{A \delTwo{p}\addTwo{Philosophical} \delTwo{a}\addTwo{Aside} on \delTwo{t}\addTwo{Time} \delTwo{i}\addTwo{Invariance}}}
\add{
Much of signal-processing theory is preoccupied with time-invariant measures, but the matter is usually presented as an assumption about the signal, and a rather restrictive, too-often unrealistic one, at that.}\add{
It can be motivated in a different way: in many settings the origin of the time scale has no intrinsic bearing on the signal and might be chosen arbitrarily.}\add{
If the observer lacks any other information by which to meaningfully situate the signal in time, such as the time of some relevant external event, then from the observer's perspective, time is subject to some large random shift, reflecting the arbitrariness of the origin.}
The question becomes, what properties of the signal can such an observer measure?\add{
As illustrated in Eq. (\ref{EQ_spectexp}), a lack of timing information makes it possible to observe only time-invariant moments.}\add{
Even if essential properties of the signal are not strictly constant over time, in the perspective of such an observer, the moment expectations  become those of a stationary signal, whose bounded statistical properties do remain constant.}
Time-shift invariant properties are particularly relevant because uncertainty of timing is, more often than not, inherent in the problem of signal identification.

\add{
One may arrive at a similar idea by noticing that the time-dependent moment of order $k$ considered at the outset in Eq. (\ref{EQ_moment}) can be trivially equated to a time-invariant moment of order $k+1$ which includes a cross-term dependence on the impulse function, $\delta(t)$; this is so because the spectrum of the latter is constant and unit valued, $\tilde{\delta}(\omega)=1$, allowing it to partake implicitly in any product.
``Situating a signal in time'' therefore amounts to observing the dependence between the signal and some time-anchoring event, encoded as an impulse.
But nothing gives such a function special importance next to any other function of doubtful relevance to the signal.  
One may conclude that time-invariant moments give a completely general description of the signal-generating process when the full structure of dependence is taken into account, whether or not the full structure includes \delTwo{any}\addTwo{signals encoding} such ``time-anchoring'' event(s). } 

\add{
In this view, one sense of what is conventionally meant by ``stationary''  becomes a statement about dependence between the signal and particular time-anchoring events.}\add{
There is another commonly used sense, which refers to whether low-order statistics, especially 2nd, vary over time.
Such fluctuations may however be described with higher-order time-invariant statistics and so do not technically require signal statistics to change over time.
For example, the tendency for signal power to fluctuate over time at a particular modulation frequency may be captured by 4th-order statistics, which encompass the power spectrum of \addTwo{bands within} the WVD.
What about the case when a glance at the spectrogram makes it plainly obvious that the signal is not stationary?
In such cases, one's eye has situated the signal in time according to some clearly identifiable series of events within the signal itself. 
But this is also, in a sense, what time-shift invariant statistics do: they take the signal itself as its own ``time-anchoring'' input. For example, in the case of PAC, phase at one frequency defines the time window over which power is observed to vary at another.
In this way, time-shift invariant 3rd-order statistics may account for a particular form of 2nd-order non-stationarity. }\add{

One important qualification should be mentioned.
It has been noted more than once now that moment estimation assumes the distribution to be bounded at the corresponding order, but it is possible to construct distributions whose moments are unbounded at or above some order (a property of Student's t-distributions, for example); thus ``stationary'' might be applied more generally to signals arising from distributions with bounded moments, while signals with unbounded moments are properly non-stationary at the given order. 
}

\add{
Perhaps it is conceptually helpful that in the spectral domain, time invariance relates to a particular subset of moment products for which time shifts cancel, making it more natural to think of the measurement of time-invariant spectral moment products as something which does not prejudge the nature of the signal.
That is, one does not need to make assumptions about time-dependent moments to measure time-invariant moments, any more than measuring a mean requires an assumption about variance. 
This view of the matter may serve to unclutter the question of stationarity, making it about the boundedness of the distribution on the one hand, and dependence on particular events on the other.
}

\section{The Bispectrum}
\label{SEC_bispectrum}
\add{
 
 \delTwo{Section 2}
\addTwo{The previous section }laid out some general context and motivation for appealing to higher-order spectra in signal processing.
The remainder of the work is concerned with addressing the two stumbling blocks identified earlier in relation to third-order statistics: computation and interpretation.
The appendices contain a more technically oriented overview of the bispectrum in which the question of how to design estimators is given particular attention.
To address the problem of interpretation, we broaden the range of estimators beyond what is normally considered.
This extension will allow us to link one class of non-standard bispectral estimator to common measures of phase-amplitude coupling.  
The following section summarizes the main results presented in \ref{section:bispectral_properties} and \ref{section:bispectral_estimators}.  

\subsection{Definition}
\label{section:definition}
Let $X$ be a random realization of a harmonizable time-series process, obeying usual assumptions of boundedness and integrability, with moments defined up to order 3.  In most cases, it will be assumed that $X$ is real-valued, unless otherwise specified. Frequency-domain representations will be indicated with a tilde; for example, $\tilde{h}(\omega)$ is the Fourier transform of $h(t)$. Scaling and normalizing constants, such as $1/2\pi$ in the inverse Fourier transform, will also be suppressed for notational  economy where doing is not expected to create confusion.

The bispectrum of the process generating $X$ is given by the following expectation in the frequency domain:
\begin{align}
\label{EQ_bispect}
	 B(\omega_1,\omega_2) =  \mathrm{E}\left[ \tilde{X}(\omega_1)\tilde{X}(\omega_2)\tilde{X}^*(\omega_1+\omega_2)\right]
\end{align}
Some insight into the meaning of this quantity comes \del{from}\add{by introducing} the inverse Fourier expansion of $\tilde{X}$ \add{into Eq. (\ref{EQ_bispect})}:
\begin{align}
\begin{split}
\label{EQ_bispect2}
	 B =  \mathrm{E}\left[
	 		\iiint{ X(r)X(s)X(t)e^{-i\omega_1(r-t)-i\omega_2(s-t)}\mathop{dr}\mathop{ds}\mathop{dt}
			} \right]\\
		  =   
	 		\iiint{\mathrm{E}\left[ X(\tau_1+t)X(\tau_2 + t)X(t)\right]e^{-i\omega_1\tau_1-i\omega_2\tau_2}\mathop{d\tau_1}\mathop{d\tau_2}\mathop{dt}
			} 
\end{split}
\end{align}
which integrates a two-dimensional Fourier transform over time.  Directly paralleling the ordinary power spectrum \citep{Cohen1989}, the integrand here may be understood as the 3rd-order Wigner-Ville distribution, up to a time-centering change of variables  \citep{gerr88,hanssen2003theory,fonoliosa1993wigner}:
\begin{align}
\label{EQ_ambi3rd}
\begin{split}
	W_3&\left(\omega_1,\omega_2,t\right)=\\&\iint X\left(\frac23\tau_1-\frac13\tau_2+t\right)X\left(\frac23\tau_2-\frac13\tau_1+t\right)X\left(t-\frac13\left(\tau_1+\tau_2\right)\right)e^{-i\omega_1\tau_1-i\omega_2\tau_2}d\tau_1\hspace{0.25em}d\tau_2
\end{split}
\end{align} 
For a third-order stationary process, the expectation of the third-order moment depends only on the relative lags of the times:
\begin{align}
\label{EQ_3rdm}
\mathrm{E}\left[X(\tau_1+t)X(\tau_2+t)X(t)\right] = \mu_3(\tau_1,\tau_2)
\end{align}
in which case, \add{substituting (\ref{EQ_3rdm}) into Eq. (\ref{EQ_bispect2})}
\begin{align}
\begin{split}
\label{EQ_bispect3}
	 B &=  \iint{ \mu_3(\tau_1,\tau_2)e^{-i\omega_1\tau_1-i\omega_2\tau_2}\mathop{d\tau_1}\mathop{d\tau_2}
			}
\end{split}
\end{align}
In other words, for a third-order stationary process, $B$ is the two-dimensional Fourier transform of the third moment, $B = \tilde{\mu}_3(\omega_1,\omega_2)$. This relationship parallels the equivalence between the power spectrum and the second moment (i.e. auto-correlation), and extends likewise to higher polyspectra and moments \citep{nikias1993signal,hanssen2003theory}. 
For a more review of the main properties of the bispectrum, the reader is invited to spend time in \ref{section:bispectral_properties}.
Matters related to the estimation of the bispectrum and, in particular, the design of bispectral estimators are tackled in \ref{section:bispectral_estimators} 

}

\subsection {\add{The Quick Summary}} 
\label{section:quick_summary}

Of the properties reviewed in \ref{section:definition} and \ref{section:bispectral_properties}, a particularly useful set for our purposes are those related to convolution.
\add{Just as the convolution of two signals in time becomes the multiplication of their spectra in the frequency domain, the bispectrum of the convolution of two signals is the product of the separate bispectra---provided the signals are statistically independent or deterministic.
Because this property allows the result to be analyzed in terms of the separate bispectra of the convolved signals, it will be useful for developing a model of recurring transient features whose timing is governed by a point process (described in \ref{sec:transient_and_oscillatory_models} and \ref{section:signal_model}).   
The bispectra of these processes are separately amenable to more detailed analysis, and properties of the bispectrum of a transient feature, in particular, will be used to establish the link with PAC.}  \add{

\ref{section:bispectral_estimators} takes a closer look at the problem of estimating bispectra. 
Just as techniques for estimating power spectra which use windowed Fourier transforms can be understood as smoothing or integrating over the ordinary Wigner-Ville distribution (see Eq. \ref{EQ_WigVille}), bispectral estimators (as well as higher-order spectral estimators generally) likewise involve integrating and smoothing over higher-order Wigner distributions \citep{gerr88,fonoliosa1993wigner,hanssen2003theory}.}\add{
Differences between alternative estimators relate to properties of the windows each applies in the smoothing or integration, but all estimators reflect the same underlying quantities related to signal moments of a given order.}
\add{

Estimators that rely on windowed Fourier transforms can be formally understood as employing a bank of single-sideband filters \citep{bingham1967modern,allen1977unified,kovach2016demodulated}.
Taking this filter-bank perspective opens a wider range of possible estimators to consideration because filter properties might be permitted to vary across bands, whereas they remain fixed in the more conventional windowed-Fourier view.
Three different classes of estimators are considered in section \ref{appendix:DifferentFilters}, which differ according to the relative bandwidths of the three bands that participate in the bispectral product. 
The results laid out in that section are used in \ref{appendix:PhPC_proof} to show the close relationship between PAC estimators and one such class of bispectral estimator.
}

\section{Phase-Amplitude Coupling}
\label{SEC_PAC}
Phase-amplitude coupling refers to dependence between the analytic envelope of an oscillatory signal component within one band and phase within another. 
For the envelope of the first component to fluctuate at the scale of the second, the bandwidth of the first component must be at least as great as the center frequency of the second and its center frequency correspondingly higher, for which reason the first component is a ``fast oscillation'' (FO), while the second is the ``slow oscillation'' (SO). 

Measures of PAC relate the analytic signal in a band encompassing the SO to the analytic envelope from the band of the FO, treating the two as separate signals, typically using a second-order statistic such as coherence, phase-locking or weighted mean vector strength.
\del{A number of elaborations of these second-order measures have been described; for example, Tort's ``modulation index,'' computes a pseudo-entropy by treating the mean amplitude as though it were a probability distribution over phase.
These extensions improve sensitivity to more general forms of dependence, but the essential qualities of PAC measures are well represented by simpler second-order metrics.}While it is also most common to use signal amplitude in quantifying PAC, one might use signal power (squared amplitude) in place of amplitude in the same way\add{, giving phase-power coherence (PhPC)}. 
\ref{appendix:PhPC_proof} demonstrates the direct relationship between PhPC and the bispectrum. 
Because conventional measures that use amplitude are related to those that use squared amplitude by a simple scalar transformation of the filtered input signals, they reflect the signal bispectrum to a first approximation and might otherwise be interpreted as bispectral estimates computed on suitably transformed input signals. 
\del{Similarly, measures that apply alternative methods of quantifying second-order dependence, such as with Kullback-Leibler divergence or other entropy-related metric over the distribution of phase, can be translated to the bispectrum with relatively little effort. 
A detailed consideration of these alternative methods of quantifying the dependence is beyond the scope of the present work,  but the essential points described in the following sections remain valid for all measures of PAC derived from the joint distribution of analytic amplitude and phase across frequency bands.} 

\subsection{\add{Multi-modal dependence}}
\label{sec:multimodal}
\add{
The second-order measures described above are most suitable when there is a single mode in the distribution of the phase difference between the SO and FO envelope.}\add{
A number of extensions develop tests that improve sensitivity to multimodal distributions, using alternative distributional measures over phase, such as Kullback-Leibler divergence or other entropy-related measures \citep{Tort2010}.}\add{
Such alternative methods of quantifying dependence can be translated to the bispectrum with relatively little effort;}\add{
the extension involves a similar treatment of the distribution of phase within the unaveraged terms in the bispectral estimator (Eq. \ref{EQ:bispectral_sum_estimator}).}\add{
The question is briefly considered again in \ref{section:multimodal_polysp}. The important take-away message for the present purpose is that the essential argument developed in the following sections remains generally valid for measures of PAC derived from the joint distribution of analytic amplitude and phase in different frequency bands.}


\section{Bispectral Signatures of PAC}
\label{section:bispectral_signatures}

\ref{appendix:PhPC_proof} contains a proof that phase-power coherence is fundamentally a bispectral estimator, which differs from conventional bispectral estimators only in the shape of the associated smoothing kernel.  
While past authors have noted some similarities between bispectral and PAC measures \citep{kramer2008sharp,hyafil2015misidentifications}, this formal relationship appears not to have been  previously described.\del{
Next we consider how the bispectrum might be applied towards identifying PAC, followed by a discussion of some broader implications for both measures.}\add{
The following sections extend this observation with a signal model that will offer some more practical insight into how the bispectrum relates to PAC.}

\begin{figure}
\centering
\includegraphics[width=120mm]{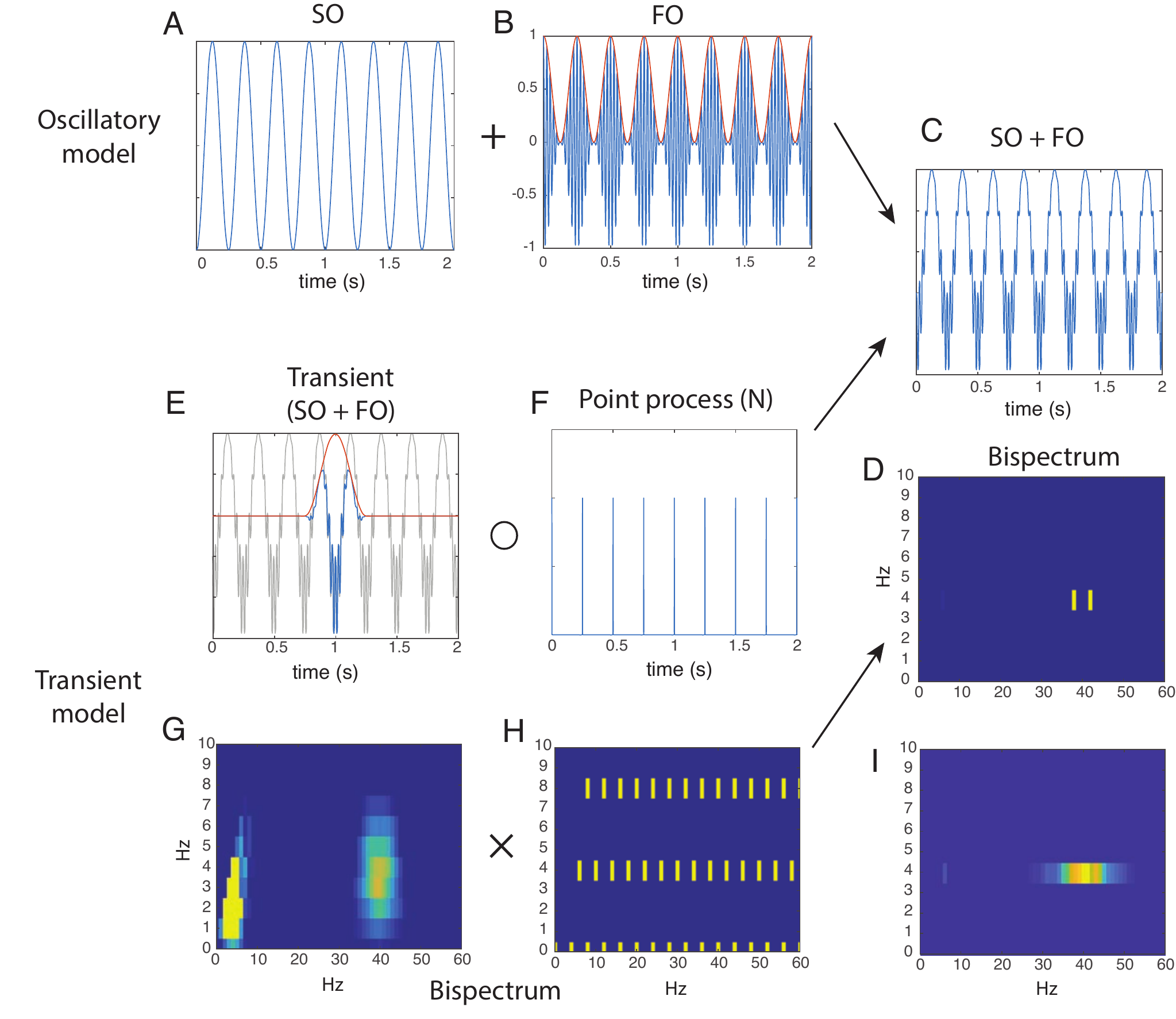}
\caption{
\label{fig:transient_oscillatory}
\add{Transient versus oscillatory models of phase-amplitude coupling. The sum of a slow oscillation (SO, \emph{A}) and a sinusoidally modulated fast oscillation (FO, \emph{B}) with modulation frequency the same as the SO, yields a signal (\emph{C}) whose bispectrum contains two peaks (\emph{D}), one at the FO frequency and another at the sum of FO and SO frequencies. 
The same signal can be modeled as the convolution of a transient SO and transiently modulated FO (\emph{E, blue line}) with a periodic train of impulse functions (\emph{F}).
Using the convolution property (described in \ref{sec:convolution}), the bispectrum of the SO+FO signal can be understood as the product of the bispectrum of the transient (\emph{G}) and that of the point process which governs its occurrence in the signal (\emph{H}).
The bispectrum of the latter contains energy only at harmonically related peaks, which restricts the comparatively smooth and broad bispectrum of the transient FO+SO to the subset of peaks at which energy in the two bispectra overlaps. 
\addTwo{In this example, the fact that the phase of the FO is fixed relative to the SO implies that the former must be an integer multiple of the latter, which is not a necessary restriction of either model. 
When the phase of the FO's are random and independent of each other and the SO, the bispectrum is no longer restricted to harmonic peaks in the range of the FO (\emph{I}).}
}}
\end{figure}

 \subsection{\add{Transient and Oscillatory Models}}
 \label{sec:transient_and_oscillatory_models}
 \add{
A case traditionally used to illustrate PAC is that of a sinusoidally modulated FO added to a sinusoidal SO of the same frequency as the FO modulating window \addTwo{(see Figure \ref{fig:transient_oscillatory})}. 
Here, the SO is a pure sinusoid,  $\cos(\theta t)$, and the FO is the sinusoidally modulated sinusoid, $(1-\cos(\theta t))\cos(\gamma t)$ . 
It is easy to get a handle on the bispectrum of this signal because we need only worry about 4 quantities (ignoring the negative half of the spectrum, which is symmetric)): peaks at $\omega=\theta$, $\omega=\gamma-\theta$, $\omega=\gamma$ and $\omega=\gamma+\theta$:
\begin{align}
\label{EQ_PureAmpMod}
\tilde{X}(\omega)=\delta(\omega-\theta) + \delta(\omega-\gamma) + \delta(\omega-\gamma-\theta)/2+ \delta(\omega-\gamma+\theta)/2
\end{align} }\add{
The bispectrum contains two peaks in the \delTwo{principle}\addTwo{principal} domain where the product $\tilde{X}(\omega_1)\tilde{X}(\omega_2)\tilde{X}^\addTwo{*}(\omega_1+\omega_2)$ does not vanish, at  $(\omega_1=\theta,\omega_2=\gamma-\theta)$ and $(\omega_1=\theta,\omega_2=\gamma)$. 
From this, it is clear that it conveys something about phase-amplitude coupling.}\add{

It is natural to try to understand spectral measures in terms of the sinusoidal functions that are elemental to spectral decompositions;
but this example offers only modest insight into what the bispectrum means in more realistic settings with signals that may be spectrally broad. 
Part of the problem is that the bispectrum, by definition, relates to the interaction between the separate components in this example.
But this means that one can't rely on any conceptually helpful decomposition according to the FO and SO considered separately to build up insight into more complicated cases.

A different tack might bring us closer to such a decomposition.}\add{
We have already noted that the spectrum of the convolution of two signals is the product of their separate spectra, a fact that generalizes to higher order spectra, provided the signals are deterministic or statistically independent at the corresponding order.}\add{
We might approach the problem by considering the signal as a series of transient features, $f_i(t-\tau_i)$, whose timing is governed by a point process, $N$.}\add{
This view of the problem decomposes the signal into a series of local features convolved with impulses generated by $N$, which encode global signal properties. 
Details of this model are worked out in \ref{section:signal_model}.}\add{
For our purpose, we will want to consider transient functions that can be decomposed into two parts, a slowly varying low-frequency part, $f_{\mathrm{SO}}(t)$ and a rapidly modulated high-frequency part $f_{\mathrm{FO}}(t)$, with $f(t) = f_{\mathrm{FO}}+f_{\mathrm{SO}}$ \addTwo{where both SO and FO are transient}.}\add{
\delTwo{Figure }\ref{fig:transient_oscillatory} \delTwo{illustrates the application of this idea to the example of sinusoidal PAC considered above.}
\addTwo{As illustrated in Figure \ref{fig:transient_oscillatory}, the transient model explains the two peaks in the bispectrum of the SO+FO signal as the restriction of the bispectrum of $f$ to peaks in the bispectrum of $N$ through multiplication.}}\add{

\subsubsection{A \delTwo{caveat}\addTwo{Caveat}}
The attentive reader may have noticed a problem with Figure \ref{fig:transient_oscillatory}, which gives the occasion to point out some limitations of the transient model:
it illustrates the case when $f$ is fixed, meaning that the phase of the FO and SO are always in the same relation and by extension that the frequency of the FO must be an integer multiple of the SO, a rather severe restriction which does not arise in the oscillatory view of the same problem.  
Fortunately, the transient model is not bound to this assumption, and a key result described in section \ref{section:outside_criterion} is that the bispectrum retains energy in the SO $\times$ FO range of frequencies even if the phase of $f_\addTwo{\mathrm{FO}}$ is random.
The limitation of the analysis presented there relates instead to the fact that it takes as a simplifying assumption that N and emitted transients, the $f_i$'s, are mutually independent.

Among the results described in \ref{section:outside_criterion}, when the phase of the $f_\addTwo{\mathrm{FO}}$'s are random or vary uniformly, the contribution of the point process to the bispectrum drops out along  a given dimension in the FO frequency range, leaving only the contribution of the transient.
In such cases, the analysis concludes that the bispectrum will still be confined to peaks in the spectrum of $N$ in the range of the SO but will not be likewise restricted in the range of the FO.}\add{
To arrive at the fact that there are only two peaks in the bispectrum when $\gamma$ is anything but an integer multiple of $\theta$ (causing the relative phase of the $f_\addTwo{\mathrm{FO}}$'s to vary), it is necessary to account for the dependence between adjacent $f_\addTwo{\mathrm{FO}}$'s, which is something the analysis neglects.}\add{
Nevertheless, as a conceptual tool it gets the general picture right, and as an analytic tool, the assumption of sequential independence is, if not always benign, at least one of the more commonly accepted transgressions in such settings. }

\subsection{The Transient Bispectrum}
\label{section:transientBispect}
\begin{figure} 
\centering
\includegraphics[width=150mm]{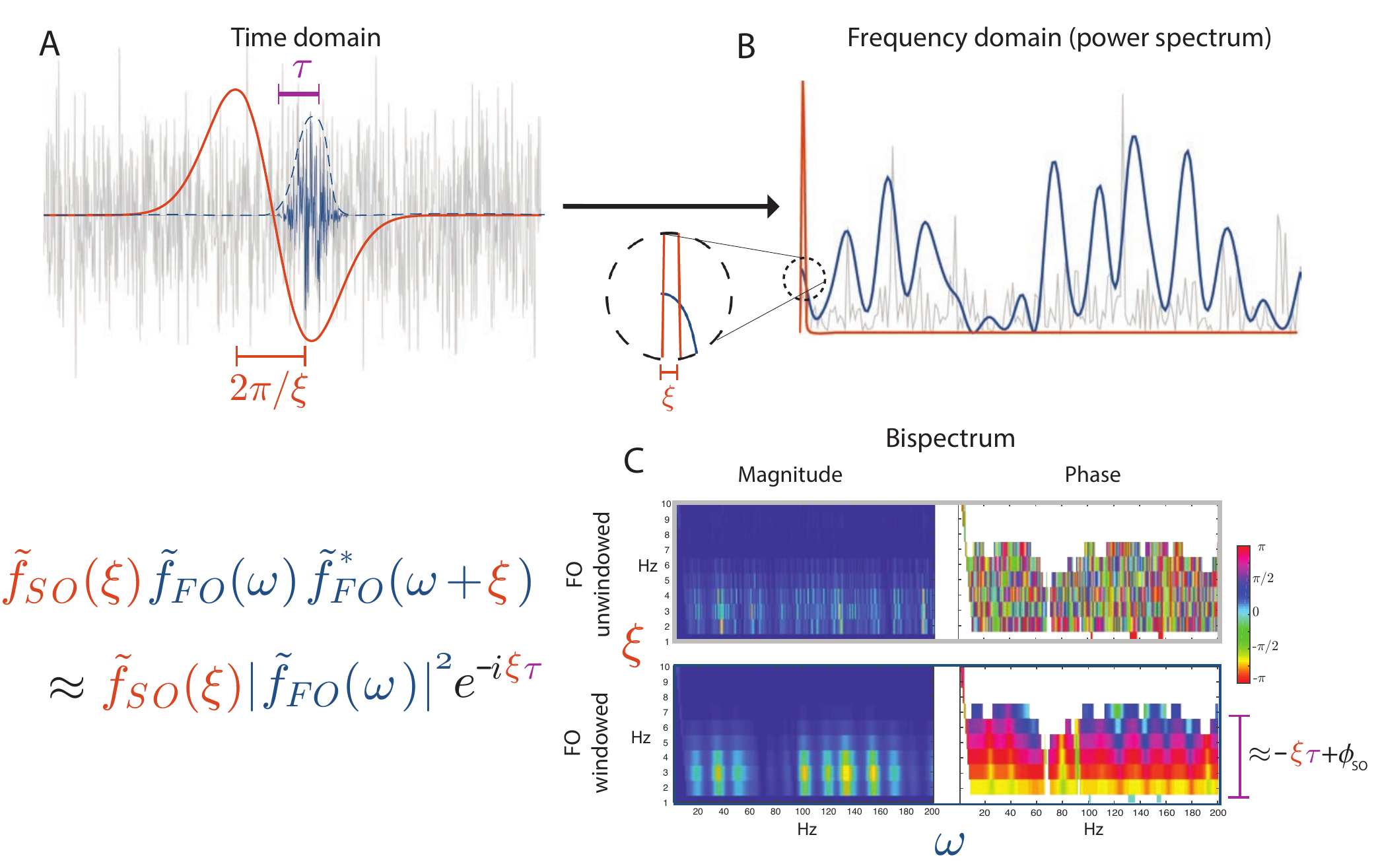}
\caption{\label{fig:FreqSmooth} 
\add{Transient modulation and spectral smoothing. \emph{Panel A:} A signal is composed of a broadband ``fast oscillation'' ($\mathrm{FO}$, \emph{\delTwo{dark }blue}) and a low frequency ``slow oscillation'' ($\mathrm{SO}$, \emph{red}). The $\mathrm{FO}$ is here depicted as white noise (\emph{\delTwo{light blue}\addTwo{gray}}) modulated by a brief window (\emph{dashed line}) whose duration is less than the characteristic time scale of the $\mathrm{SO}$, $2\pi\xi^{-1}$. 
\emph{B}: In the frequency domain, such transient windowing forces the spectrum of the $\mathrm{FO}$ to be smooth at the frequency scale of $\xi$ (\emph{inset}) so that the product $\tilde{f}_{\mathrm{FO}}(\omega_1)\tilde{f}^*_{\mathrm{FO}}(\omega_1+\xi)$  may be locally approximated by the power spectrum of the $\mathrm{FO}$ multiplied by a complex sinusoid which encodes the delay of the window function, $e^{-i\xi\tau}$ (see Eq. \ref{EQ_PACapprox}). Contrast the unsmoothed power spectrum in the absence of time modulation, (\emph{\delTwo{light blue}\addTwo{gray}}). \emph{C:} As a result of this relationship, the bispectrum of the composite signal contains information about the spectrum of the $\mathrm{SO}$ and power spectrum of the $\mathrm{FO}$ \addTwo{. The relative delay between the $\mathrm{FO}$ and $\mathrm{SO}$ is encoded in bispectral phase (\emph{C, lower right}), which is approximately linear in the range of the $\mathrm{SO}$; contrast the random phase of the unwindowed FO (\emph{C, upper right})}
}}
\end{figure}

\begin{figure} 
\centering
\includegraphics[width=40mm]{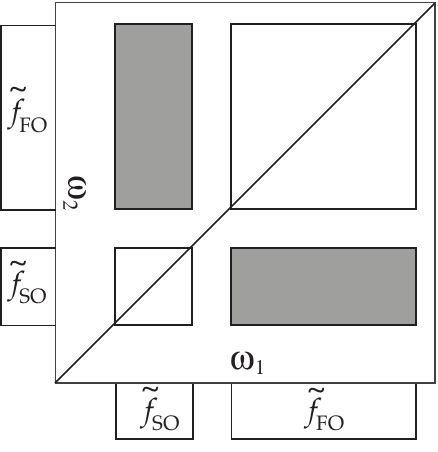}
\caption{\label{fig:Fregions} 
Regions of the bispectrum used by the inside and outside criteria. Phase-amplitude coupling is characterized by  energy in the ``outside'' regions given by the support of $\tilde{f}_{\mathrm{SO}}\tilde{f}_{\mathrm{FO}}$ (\emph{gray boxes}) while \del{harmonic} features related to \del{signal transients} \add{sharp peaks and edges} also tend to create energy in the ``inside'' regions (\emph{white boxes}).}
\end{figure}

\add{The main advantage of the transient model and its relevance for understanding PAC is explained next.}
If the mean of $f$, $\tilde{f}(0)$, is nonzero, it is easily seen that the bispectrum of $f$ contains its own power spectrum, $\tilde{f}(0)\left|\tilde{f}(\omega)\right|^2$, along the axes $\omega_1=0$ and $\omega_2=0$.
\add{This observation can be extended more generally to slow oscillations beyond the DC axis:} suppose $f$ is the sum of two components, $f=f_{\mathrm{SO}}+f_{\mathrm{FO}}$, the first of which, $f_{\mathrm{SO}}$, occupies a narrower bandwidth than the second at center frequency, $\xi$,  while the amplitude of the second rises and falls transiently at a time scale shorter than the period of the first. 
This modulation  implies that $f_{\mathrm{FO}}$ is effectively windowed at a time scale less than $2\pi\xi^{-1}$. 
\add{The consequence of such time-domain windowing on the frequency domain mirrors exactly the effect of frequency-domain windowing on the time domain (i.e. filtering), which is to say the result is a smoothed version of the unwindowed spectrum. 
In particular, windowing at the scale $2\pi\xi^{-1}$ implies that the spectrum of $f_{\mathrm{FO}}$ is smooth at the scale of $\xi$,} justifying the following approximation
\begin{align}
\label{EQ_f2smooth}
\tilde{f}_{\mathrm{FO}}(\omega ) \tilde{f}^*_{\mathrm{FO}}(\omega + \xi) \approx \left|\tilde{f}_{\mathrm{FO}}(\omega)\right|^2e^{-i\xi\tau}
\end{align}
where $\tau$ accounts for some arbitrary time delay.  
These points are illustrated in Figure \ref{fig:FreqSmooth}. 
The term $e^{-i\xi\tau}$ describes the phase of the amplitude modulation relative to the emission time. 
We may partition the bispectral plane according to the spectral ranges of $f_{\mathrm{SO}}$ and $f_{\mathrm{FO}}$, as shown in Fig. \ref{fig:Fregions}.
Where energy falls within this partitioning gives the first set of criteria for distinguishing PAC.

\subsection{Bispectral Definition of PAC}
 \label{section:bispectral_definition_of_pac}
\begin{figure} 
\centering
\includegraphics[width=150mm]{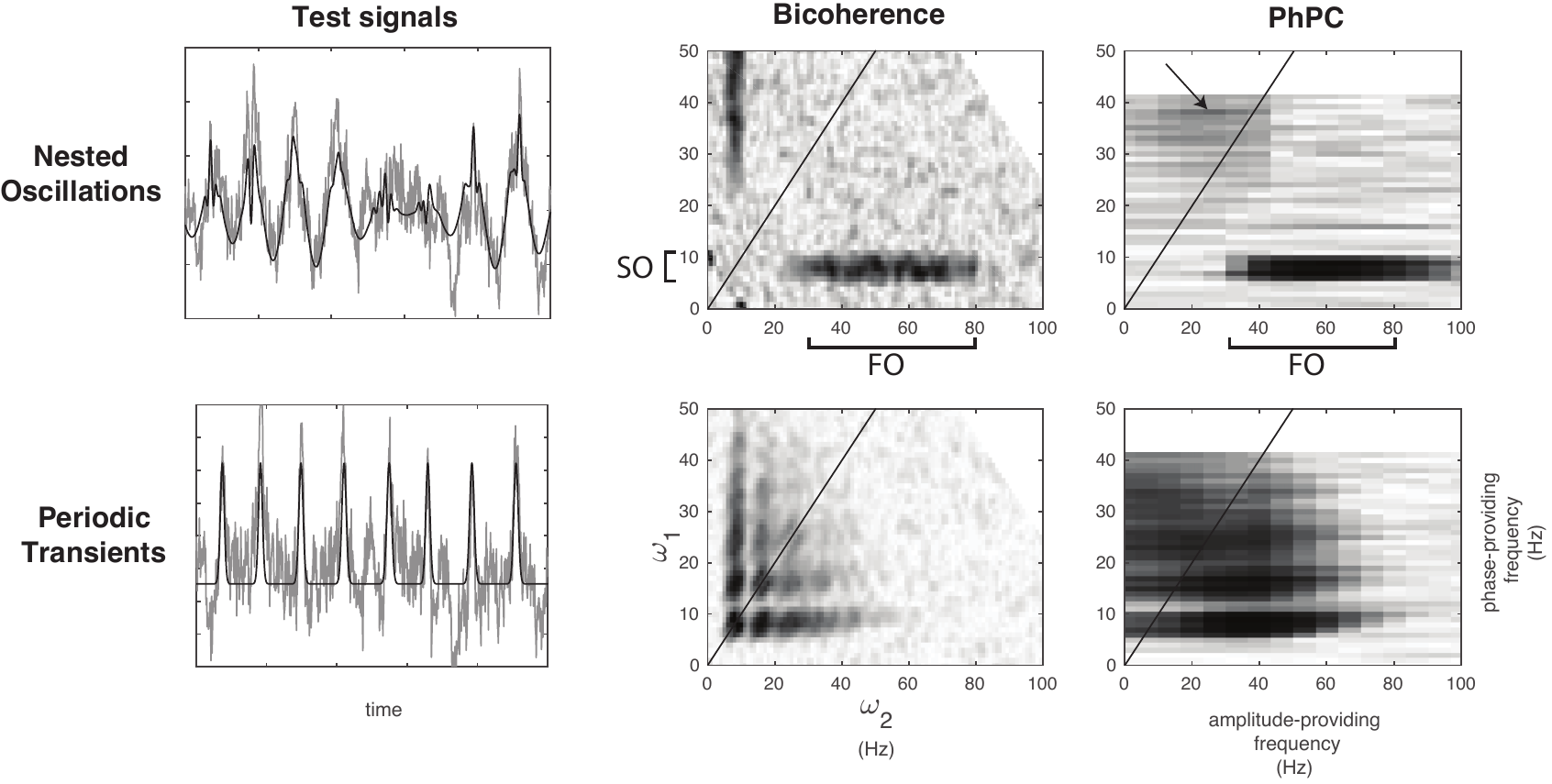}
\caption{\label{fig:test_signals} 
Comparison of bicoherence and phase-power coherence, a PAC measure, for two test signals, one modeling a periodic broad-spectrum transient and the other, a nested oscillation with random phase in the FO.  
\emph{Top Row}: The first test signal was designed to contain nested oscillations by adding 30-80 Hz filtered white noise amplitude modulated according to the phase in a second 6-10 Hz band of filtered noise, which was also added, along with a background of 0 dB $1/f$ noise (\emph{top left panel}; signal with noise is \emph{gray line}). 
\emph{Bottom Row}: For the second test signal, a periodic train of transients was simulated as $\exp{[10\cos\left(\phi(t)\right)]}$ where $\phi(t)$ is analytic phase from 6-10Hz filtered white noise, to which 0 dB $1/f$ noise was added (\emph{bottom left panel}). 
\emph{Middle Column}: Bicoherence clearly reveals the presence of nested oscillations in the first test signal (\emph{top middle panel}). 
For the second test signal, the origin of cross-frequency coupling in sharp transients is reflected by harmonic structure and terms on the diagonal (\emph{bottom middle panel}). 
\emph{Right Column}: Phase-power coherence computed with 1 Hz bandwidth for the phase-providing band and 40 Hz for the amplitude-providing band also reveals the presence of nested oscillations in the first test signal (\emph{top right panel}). 
Anisotropic smoothing in measures of PAC obscures the symmetry of the bispectrum under exchange of axes, but  the effect of symmetric terms can still be discerned as a slight increase in coupling between low frequency amplitude high-frequency phase (\emph{arrow}). 
For the second test signal, the harmonic structure is reflected in the bands within the phase-providing frequencies, but the relative importance of diagonal terms is obscured by smoothing along the amplitude-frequency ($\omega_2$) axis (\emph{bottom right panel}). 
 }
\end{figure}

First, we need a more precise statement of what is meant by PAC.
An often-cited example of what is \emph{not} meant are cases of ``spurious'' PAC that arise from recurring transient signal features, such as spectrally broad ``spikes'' or sharp-edged waves, which may or may not be periodic (see Figure \ref{fig:test_signals}) \citep{kramer2008sharp,lozano2016neuronal}. 
Such false PAC will tend to exhibit a consistent phase relationship across a range of frequencies, in particular, along the harmonics of  any fundamental periodicity.
``True'' PAC is taken here to mean nested oscillations with the following characteristics: 
\begin{enumerate}
\item{For a fast oscillation (FO) to be ``nested,'' within a slow oscillation (SO), its amplitude must vary at a time scale around the period of the slow oscillation.}
\item{The FO should be concurrent with the SO.}
\item{The phase of the FO must either lack any consistent phase relationship with that of the SO or it must fall within a band that is isolated from the \del{surrounding harmonics of the FO}SO and any related harmonics. }
\end{enumerate}

The first point implies that each burst of the FO occupies a bandwidth wider than the center frequency of the slow oscillation in which it is embedded, giving a characteristically smooth spectrum at the scale of the SO. 
The second point excludes non-concurrent responses; for example,  a fast oscillation followed by a slower one. 
The third point excludes spectrally broad features associated with sharp transients and ensures that any oscillation nested in the SO with a consistent phase relationship is genuinely oscillatory.
These points justify \delTwo{bispectral criteria}\addTwo{guidelines} for identifying PAC \addTwo{in the bispectrum}, outlined in the following sections.
More generally, they provide \delTwo{an entry point for characterizing}\addTwo{a way to characterize} recurring signal features through the bispectrum.

\subsubsection{Defining PAC: The Outside Criterion}
\label{section:outside_criterion}
The first criterion for PAC is that the time scale over which the amplitude of the nested fast oscillation,  $f_{\mathrm{FO}}$, varies must be less than the period of the slow oscillation, $f_{\mathrm{SO}}$, in which it is embedded. 
When this \del{condition applies, and}\add{holds, the approximation in Eq. (\ref{EQ_f2smooth}) implies that} the power spectrum of $f_{\mathrm{FO}}$ appears \addTwo{in the bispectrum} parallel to the $\omega_2$ axis within the support of $\tilde{f}_{\mathrm{SO}}$ because within this range
\begin{align}
\label{EQ_PACapprox}
\tilde{f}(\omega_1)\tilde{f}(\omega_2)\tilde{f}^*(\omega_1+\omega_2)\approx \tilde{f}_{\mathrm{SO}}(\omega_1)\left|\tilde{f}_{\mathrm{FO}}(\omega_2)\right|^2e^{i\omega_1\tau}
\end{align}
and likewise for the $\omega_1$ axis under symmetry\del{, where $\tau$ is the time delay}\add{. The time delay, $\tau$, recovers the relative lag} between $f_{\mathrm{SO}}$ and \addTwo{the envelope} of $f_{\mathrm{FO}}$.
It can be seen that the power spectrum of $f_{\mathrm{FO}}$ is reproduced parallel to the $\omega_1$ axis at $\xi$, while in the orthogonal direction, the spectrum of $f_{\mathrm{SO}}$ is reproduced parallel to $\omega_2$ within the support of $\tilde{f}_{\mathrm{FO}}(\omega_1)$.  \add{}
Examples of this effect can be found in Figure \ref{fig:test_signals} (middle panels) for both the nested-oscillation and transient test signal. 

Applying a similar analysis to the stochastic case describe in \ref{section:random_features}, suppose that the spectrum of $f_{\mathrm{FO}}$ overlaps negligibly with $f_{\mathrm{SO}}$, so that  $\tilde{f}_{\mathrm{SO}i}(\omega)\tilde{f}_{\mathrm{FO}j}(\omega)\approx 0$, then within the support of $\tilde{f}_{\mathrm{SO}}(\omega_1)\tilde{f}_{\mathrm{FO}}(\omega_2)$ we are left with
\begin{align}
\label{EQ_PPBC3}
\begin{split}
			 \mathrm{E}\left[\tilde{X}(\omega_1)\tilde{X}(\omega_2)\tilde{X}^*(\omega_1+\omega_2)\right]&=\\
			  &
			\left<\tilde{f}_{\mathrm{SO}}(\omega_1)\tilde{f}_{\mathrm{FO}}(\omega_2)\tilde{f}_{\mathrm{FO}}^*(\omega_1+\omega_2)\right>\\
			 &+ \left<\tilde{f}_{\mathrm{SO}}(\omega_1)\right>
			 \left<\tilde{f}_{\mathrm{FO}}(\omega_2)\tilde{f}_{\mathrm{FO}}^*(\omega_1+\omega_2)\right>
			  \tilde{\mu}_2(\omega_1)\\
			  &+ \left<\tilde{f}_{\mathrm{FO}}(\omega_2)\right>
			 \left<\tilde{f}_{\mathrm{SO}}(\omega_1)\tilde{f}_{\mathrm{FO}}^*(\omega_1+\omega_2)\right>
			  \tilde{\mu}_2(\omega_2)\\
			 &+ \left<\tilde{f}_{\mathrm{SO}}(\omega_1)\tilde{f}_{\mathrm{FO}}(\omega_2)\right>
			 \left<\tilde{f}_{\mathrm{FO}}^*(\omega_1+\omega_2)\right>
			 \tilde{\mu}_2(\omega_1+\omega_2)\\
			 &+ \left<\tilde{f}_{\mathrm{SO}}(\omega_1)\right>\left<\tilde{f}_{\mathrm{FO}}(\omega_2)\right>\left<\tilde{f}_{\mathrm{FO}}^*(\omega_1+\omega_2)\right> \tilde{\mu}_3(\omega_1,\omega_2)\\
\end{split}
\end{align}
\del{When }\deleq{$f_{\mathrm{SO}}$}\del{ and }\deleq{$f_{\mathrm{FO}}$}\del{ are deterministic this reduces to Eq. }\deleq{(\ref{EQ_PPBC})}\del{. At the other extreme, s}Suppose the phase of $f_{\mathrm{FO}}$ is random such that $\left<\tilde{f}_{\mathrm{FO}}\right>=\left<\tilde{f}_{\mathrm{FO}}\tilde{f}_{\mathrm{SO}}\right>=0$, then only the first two terms of Eq. (\ref{EQ_PPBC3}) remain: 
\begin{align}
\label{EQ_PPBC4}
\begin{split}
			 \mathrm{E}&\left[\tilde{X}(\omega_1)\tilde{X}(\omega_2)\tilde{X}^*(\omega_1+\omega_2)\right]\\			  
			&=\left<\tilde{f}_{\mathrm{SO}}(\omega_1)\tilde{f}_{\mathrm{FO}}(\omega_2)\tilde{f}_{\mathrm{FO}}^*(\omega_1+\omega_2)\right>
			 + \left<\tilde{f}_{\mathrm{SO}}(\omega_1)\right>
			 \left<\tilde{f}_{\mathrm{FO}}(\omega_2)\tilde{f}_{\mathrm{FO}}^*(\omega_1+\omega_2)\right>
			  \tilde{\mu}_2(\omega_1)\\
			  &\approx\left<\tilde{f}_{\mathrm{SO}}(\omega_1)\left|\tilde{f}_{\mathrm{FO}}(\omega_2)\right|^2e^{\add{-}i\omega_1\Delta\tau}\right>
			 + \left<\tilde{f}_{\mathrm{SO}}(\omega_1)\right>
			 \left<\left|\tilde{f}_{\mathrm{FO}}(\omega_2)\right|^2e^{-i\omega_1\Delta\tau}\right>
			  \tilde{\mu}_2(\omega_1)\\		  
\end{split}
\end{align}
The first term does not vanish when there is a consistent relationship of phase between $f_{\mathrm{SO}}$ and the amplitude modulation of $f_{\mathrm{FO}}$. 
The second term remains when there is also a consistent lag between both terms and the emission of the point process, which is reflected accordingly in the weighting by the spectrum of the point process, $\tilde{\mu}_2(\omega_1)$. 
But because we are free to define the emission times of the point process such that $\tau_{\mathrm{SO}}=0$, the second term merely reflects the power spectral contribution of the driving point process. 
The bispectral estimate within the support of $f_{\mathrm{SO}}(\omega_1)f_{\mathrm{FO}}(\omega_2)$ therefore depends on the timing of the amplitude of $f_{\mathrm{FO}}$ relative to the phase of $f_{\mathrm{SO}}$, but not the phase of $f_{\mathrm{FO}}$.

\begin{figure} 
\centering
\includegraphics[width=150mm]{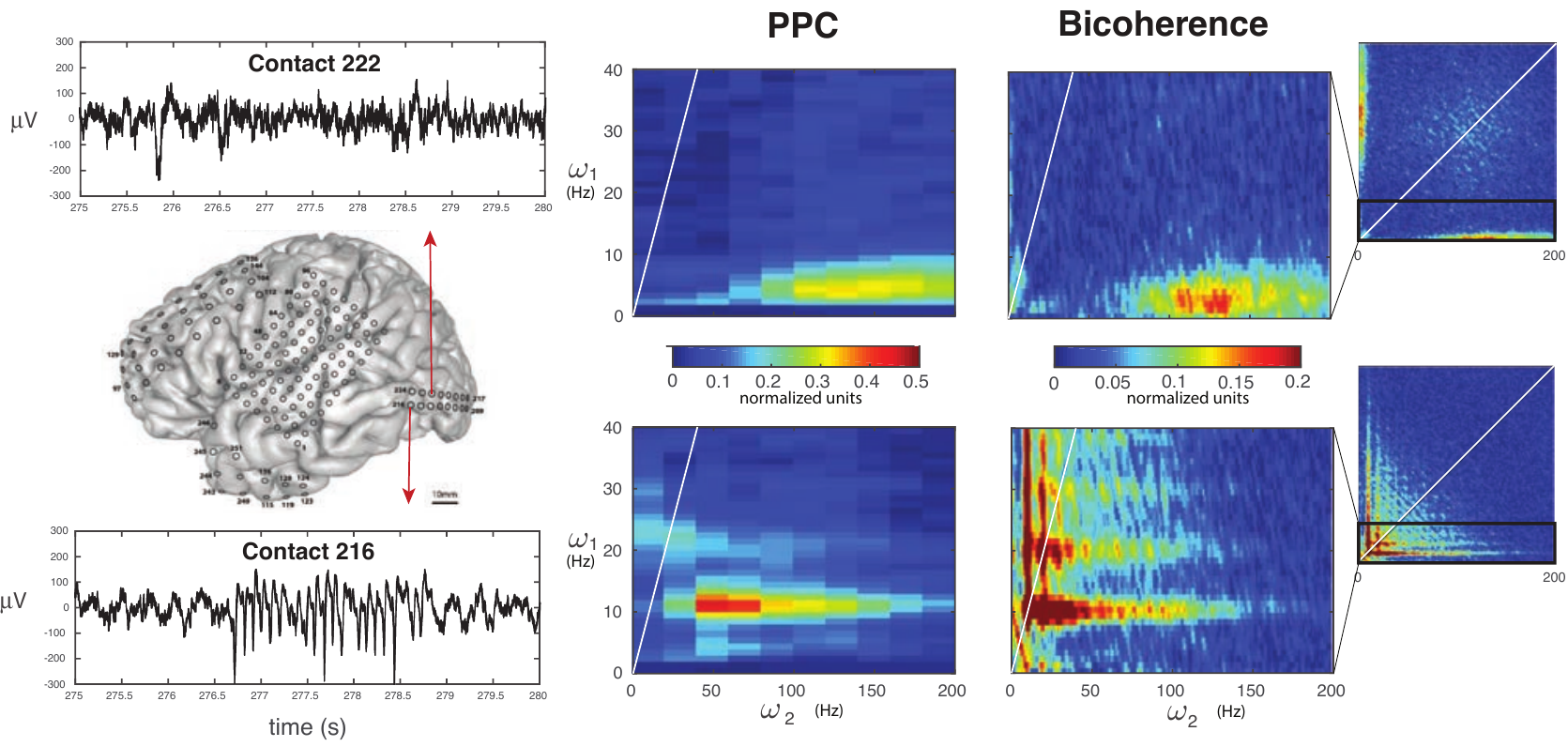}
\caption{\label{fig:275_examp} 
Distinguishing the origin of PAC in human ECoG data with bicoherence.  Twenty minutes of ECoG data were obtained from a patient undergoing invasive clinical monitoring for epilepsy while the patient watched a television show. Contact 216 over occipital cortex (\emph{bottom row}) recorded frequent interictal spikes, while the neighboring contact, 222 (\emph{top row}) contained few spikes. \emph{Left column}: A 5 second window showing a typical series of discharges in channel 216 (\emph{bottom left panel}). \emph{Middle column}: Both channels exhibit strong apparent phase-amplitude coupling as measured by phase-power coherence, in channel 216, between 50-150 Hz gamma power and 10 Hz phase (\emph{bottom middle panel}) and 100-200 Hz power and 2-5 Hz phase in channel 222 (\emph{top middle panel}). \emph{Right column}:  NNB Bicoherence reflects the phase-amplitude coupling in both cases but also reveals the presence of harmonic diagonal and off-diagonal terms in channel 216 (\emph{bottom right panel}, white line indicates $\omega_1=\omega_2$), suggesting that PAC in this case originates from the sharp-edged ictal discharges. In channel 222 (\emph{top right panel}), bicoherence reveals no harmonic structure and little energy \del{along the diagonal and off-diagonal} \add{around the transition from the inside to outside regions, implying that PAC cannot likely be explained by sharp-edged signal features. 
Energy near the center of the inside region (\emph{upper right inset}) suggests that he phase of the nested FO is not completely random with respect to its own envelope.}
Both insets show symmetric bicoherence over the full range of frequencies; regions shown in the adjoining panels are outlined in black.}
\end{figure}

This form of cross-frequency dependence fulfills sensible criteria for phase-amplitude coupling: it implies a correlation between the amplitude of the amplitude-providing signal component, $f_{\mathrm{FO}}$, and an underlying oscillation with a longer period than the time scale of the amplitude modulation of $f_{\mathrm{FO}}$, which is what is meant conventionally by phase-amplitude coupling. 
Reflecting \del{its time scale}\add{the scale of its modulation}, the spectrum of $f_{\mathrm{FO}}$ must be both relatively broad and smooth, but because the power spectrum discards phase information in $f_{\mathrm{FO}}$,  the phase of $f_{\mathrm{FO}}$ does not require any consistent relationship with $f_{\mathrm{SO}}$, which is also a frequent characteristic of nested oscillations in phase-amplitude coupling. 
This condition does not, however, rule out a contribution from spectrally broad transients (see example in Fig.  \ref{fig:test_signals}, lower middle panel), a possibility that motivates the second and third criteria, next. 

\subsubsection{Defining PAC:  The Inside Criterion}
\label{section:inside_criterion}
The central part of the bispectrum covers the  regions of support for $\tilde{f}_{\mathrm{SO}}(\omega_1)\tilde{f}_{\mathrm{SO}}(\omega_2)$ and $\tilde{f}_{\mathrm{FO}}(\omega_1)\tilde{f}_{\mathrm{FO}}(\omega_2)$ (white boxes in Fig. \ref{fig:Fregions}).  Non-vanishing terms in the bispectrum within these regions \add{\delTwo{reflect}\addTwo{relate to} the separate bispectra of the components. }\del{tend to reflect spectrally broad features associated with sharp-edged transients or harmonic structure when the driving process is periodic.}\add{
\delTwo{In the context of a SO accompanied by a transiently modulated FO, energy}\addTwo{Energy} in the $\mathrm{FO}\times\mathrm{FO}$ ``inside'' box of Fig. \ref{fig:Fregions} reflects the consistency between the phase of the FO and its \addTwo{own} modulating envelope.}\add{ 
For example, if the FO is obtained by shifting $\cos(\beta (t + \tau))$ by some random delay, $\tau$ and windowing with an unshifted envelope, $h$:}\footnote{\add{It is assumed here for the sake of argument that $h$ varies slowly relative to $2\pi\beta^{-1}$ to avoid any complications related to overlapping spread from negative and positive frequencies (i.e. Bedrosian's theorem applies) but quickly enough to allow for a non-vanishing bispectral product.}}
\add{\[
f_{\mathrm{FO}} = h(t)\cos(\beta (t + \tau))
\]}
\add{then its bispectrum is given by
\begin{align}
\label{EQ_nested}
\mathrm{E}\left[\tilde{f}_{\mathrm{FO}}(\omega_1)\tilde{f}_{\mathrm{FO}}(\omega_2)\tilde{f}^*_{\mathrm{FO}}(\omega_1+\omega_2)\right] = \tilde{h}(\omega_1-\beta)\tilde{h}(\omega_2-\beta)\tilde{h}^*(\omega_1+\omega_2-\beta)\mathrm{E}\left[e^{i\beta\tau}\right]
\end{align}}\add{
which vanishes if the variability in $\tau$ suffices to make $e^{i\beta\tau}$ uniform \delTwo{and does not otherwise}\addTwo{in the spectral range of the FO.}
Energy near the center of the inside box therefore implies a consistent relationship between the modulating envelope of the FO and its own phase. 
} 

\add{We observed similarly in \ref{section:outside_criterion} that energy in the outside box has to do with consistency between the envelope of the FO and the phase of the SO. 
Such consistency may of course, in both cases, be the result of some spectrally broad feature that we do not wish to call a nested oscillation, which will tend to manifest as consistent phase across a broad range of frequencies spanning the transition from inside to outside boxes.}\add{
One might reason that energy in both boxes implies a consistent phase relationship between the SO and the FO, but this need not always be the case.
Variability in the timing of the FO modulating window may happen on the scale of the SO without abolishing energy in the outside box, and this scale might be large relative to the frequency range of the FO.}\add{
Variability of the FO delay will, however, attenuate the bispectrum around the transition from outside to inside boxes, because any energy in the SO at frequencies above the scale of the variability will be washed out. 
This will have the effect of accentuating the separation between the inside and outside boxes as the contribution of higher-frequency components of the SO is suppressed by variability in the timing of the FO window.}\add{

One may conclude from this that FO's which are either spectrally isolated, and hence oscillatory, or those that occur with some variability of delay may generate modes in the inside region that clearly stand apart from energy in the outside region.\footnote{\add{It can also be pointed out that energy in any of the regions might be the result of entirely unrelated signal components; it is assumed for the present that the signal bispectrum is dominated by one component. The problem of decomposing multi-components signals into their independent parts is an important open question briefly considered in section \ref{sec:future}. }}
In contrast, energy generated by a sharp-edged transient will tend to exhibit a tight phase relationship among all components, leading to a continuous smearing of energy across  both regions.
In the presence of periodicity, such energy will cluster at a lattice of harmonically-related peaks along axes given by $m\omega_1=n\omega_2$ where $\tilde{f}(n\omega)\tilde{f}(m\omega) \tilde{f}^*\left((n+m)\omega\right)$ is non-vanishing \cite{Sheremet4218}.
\addTwo{This effect is a well-known source of spurious phase-amplitude coupling \cite{lozano2016neuronal}, as well as spurious $n:m$ phase locking \cite{Sheffer-Teixera16}.} }
\deleq{\del{. When the phase of }$f_{\mathrm{FO}}$\del{ remains in a consistent relationship with that of }$f_{\mathrm{SO}}$\del{ any of the third, fourth and fifth terms in Eq. (}\ref{EQ_PPBC4}\del{) may become non-vanishing; for a periodic driving process, these terms create harmonic banding within the support of }$f_{\mathrm{FO}}$\del{ and peaks at the aforementioned lattice points.} }

\add{Figure \ref{fig:test_signals} shows a comparison between such a periodic transient and nested oscillations involving a purely random FO.}
\add{Figure \ref{fig:275_examp} shows a very similar comparison for real data recorded from two neighboring regions of lateral occipital cortex in an epilepsy patient. 
Frequent inter-ictal spiking in one channel (Fig. \ref{fig:275_examp}, channel B, lower panels) results in what looks superficially like PAC between phase around 10 Hz and power in the 50-150 Hz range.}\add{
In the bispectral measure, this apparent phase-amplitude coupling is accompanied by clearly visible harmonic banding and a wide monotonically diminishing spread of energy, which is much less easy to see in the PAC measure, due to the smoothing \delTwo{properties}\addTwo{and limited field of view} of the latter.}\add{ 

In a \addTwo{different} nearby channel, both measures reveal an association between phase in the 1-4 Hz range and power in the 80-200 Hz range, in the absence of a similar broad smear of energy---this looks more like the picture expected of phase-amplitude coupling related to nested oscillations.}\add{
In the bispectral measure, some energy also appears near the center of the inside region (Fig. \ref{fig:275_examp}, inset of the upper right panel), implying that the FO exhibits some degree of consistency in its form (that is, phase is not completely random with respect to the modulating window), but the mode clearly separates from the outside region, making the attribution of PAC reasonable.
Because PAC measures do not include the inside region within their field of view, this detail is unavailable from them. 
As this example illustrates, both kinds of phenomena, PAC and sharp waves, occur as part of ongoing activity \citep{vaz2017dual}; discriminating between them is crucial to the interpretation of PAC.}

\begin{figure}
\centering
\includegraphics[width=150mm]{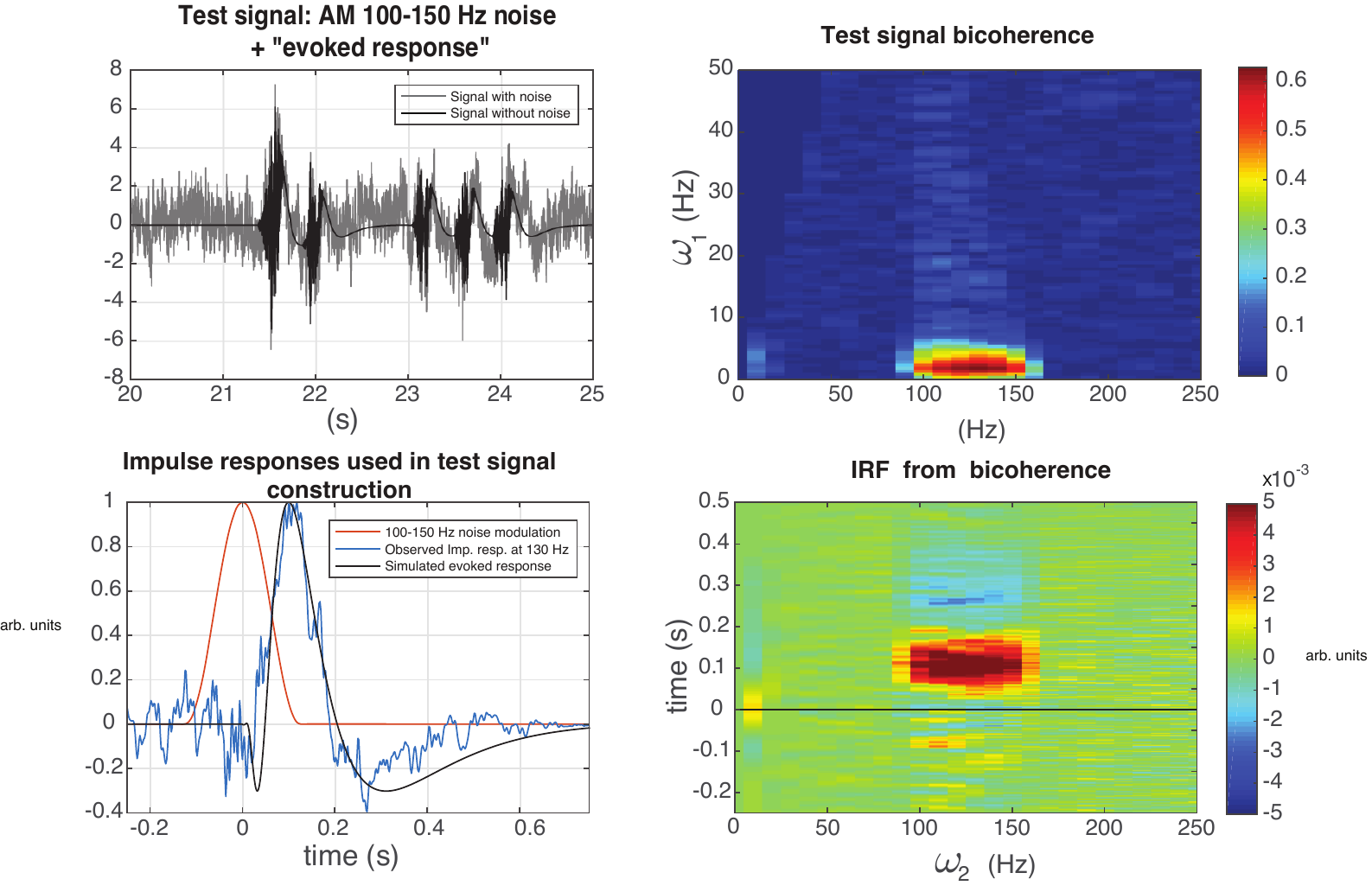}
\caption{\label{TimeDelayExamp}
Recovering signal features from the bispectrum. \emph{Top Left}: The test signal (\emph{black line}) simulates a series of high-frequency oscillations (FO) followed by a slower transient response, resembling a physiological evoked response embedded in Gaussian $1/f$ noise at random times (\emph{gray line}, signal with noise). \emph{Top Right}: Bicoherence reflects the association between the FO and the following slow response.  \emph{Bottom Row}: The \addTwo{one-dimensional inverse Fourier transform, taken along $\omega_1$, gives a function that approximates the impulse response relating the SO to modulations of power at the corresponding FO frequency (See Eq. \ref{EQ_IRF})} \delTwo{SO multiplied by the power-spectrum of the FO along }\deleq{$\omega_2$}\delTwo{  impulse response obtained from an inverse Fourier transform along}\deleq{ $\omega_1$} (\emph{Right}). \addTwo{For a transiently modulated FO, the impulse response  approximate\addTwo{s}\delTwo{ly recovers} the }\delTwo{slow response}\addTwo{SO} over the support of the FO along $\omega_2$ (\emph{Left}). 
}
\end{figure}

\subsubsection{Defining PAC: The Time-Delay Criterion}
\label{section:impulse_response}
 Eq.  (\ref{EQ_PACapprox}) implies that the bispectrum contains potentially useful information about the relative timing of  SO phase and FO amplitude modulation. Information about timing can be recovered with an inverse Fourier transform along one of the frequency dimensions, as shown in Fig. \ref{TimeDelayExamp}.
 For the homogenous point process with constant $\lambda$, in the presence of pure PAC:

\begin{align}
\label{EQ_IRF}
\begin{split}
I(\tau,\omega_2) &= \int_{SO}{B(\omega_1,\omega_2)e^{i\omega_1\tau}\mathop{d\omega_1}}\\
			      &\approx f_{\mathrm{SO}}(\tau+\Delta\tau)\left|\tilde{f}_{\mathrm{FO}}(\omega_2)\right|^2 + \tilde{f}_{\mathrm{\add{SO}}}(\omega_2)e^{\omega_2 \Delta\tau}f_{\mathrm{FO}}^{(2)}(\tau)
\end{split}
\end{align}
\add{where the integral covers the support of $f_{SO}$, excluding the diagonal symmetry region in quadrants II and IV.} Because $f_{\mathrm{SO}}$ and $f_{\mathrm{FO}}$ occupy non-overlapping spectral ranges, the result is a two-dimensional function with two bands, the first around the center frequency of $f_{\mathrm{SO}}$, containing the autocorrelation of  $f_{\mathrm{FO}}$ scaled by $\tilde{f}_{\mathrm{SO}}(\omega_2)$ and the second within the support of $\tilde{f}_{\mathrm{FO}}$, containing $f_{\mathrm{SO}}$ scaled by $\left|\tilde{f}_{\mathrm{FO}}(\omega_2)\right|^2$, both shifted by the time delay between them. 
For a periodic $N$, the terms are reduplicated at the corresponding periodicity. 

This \addTwo{time-domain representation is useful for verifying that the FO and SO overlap in time, as should be the case for nested oscillations.}\addTwo{ 
More generally it provides a way to relate PAC to specific features of SO.
It also illustrates the broader point that because the bispectrum preserves phase information, it encodes details of waveform shape that are lost in standard spectral analyses \citep{bartelt1984phase}. 
}

\section{\addTwo{Extensions}}
\label{section:Extensions}

\subsection{\addTwo{Multi-modal dependence}}
\label{section:multimodal_polysp}
\addTwo{As mentioned in section \ref{sec:multimodal}, a variety of extensions to basic measures of PAC address more complicated forms of dependence between phase and amplitude, as when peaks of amplitude occur at two or more separate phases.  
It was also noted that similar extensions might be developed for the distribution of phase of within bispectral estimators. 
A major advantage of standard bispectral estimators is that their capacity to reveal phase-amplitude dependence is not inherently limited by the analysis bandwidths, but the need to construct a distribution over phase in such extensions is bound to increase sensitivity to the parameters of the analysis filter.}

An \addTwo{alternative strategy for this problem might appeal to subregions within 4th and higher order spectra. 
For example, if the modulation of the FO is doubled in frequency compared to the SO, meaning there are two peaks of amplitude at opposite phase over every $2\pi$ cycle of phase in the SO, this might be revealed within the $\omega_1=\omega_3$ plane of the \emph{trispectrum}:
\begin{align}
\label{EQ_frdouble}
\mathrm{E}\left[\tilde{X}^2(\omega_1)\tilde{X}(\omega_2)\tilde{X}^*(2\omega_1+\omega_2)\right]
\end{align}}
\addTwo{Likewise, a distribution with $k$ peaks in the amplitude of the FO should be revealed within the $\omega_1=\omega_3=\omega_4=\dots=\omega_k$ plane of the $(k+1)^\mathrm{th}$ order spectrum:
\begin{align}
\label{EQ_frdouble}
\mathrm{E}\left[\tilde{X}^k(\omega_1)\tilde{X}(\omega_2)\tilde{X}^*(k\omega_1+\omega_2)\right]
\end{align}}
\addTwo{Because these pertain to a two-dimensional subspace in each case, the full polyspectrum does not need to be estimated to obtain a test of $k$-mode dependence, and estimation should therefore not greatly increase the computational burden over that of the bispectrum.}

\begin{figure}
\centering
\includegraphics[width=100mm]{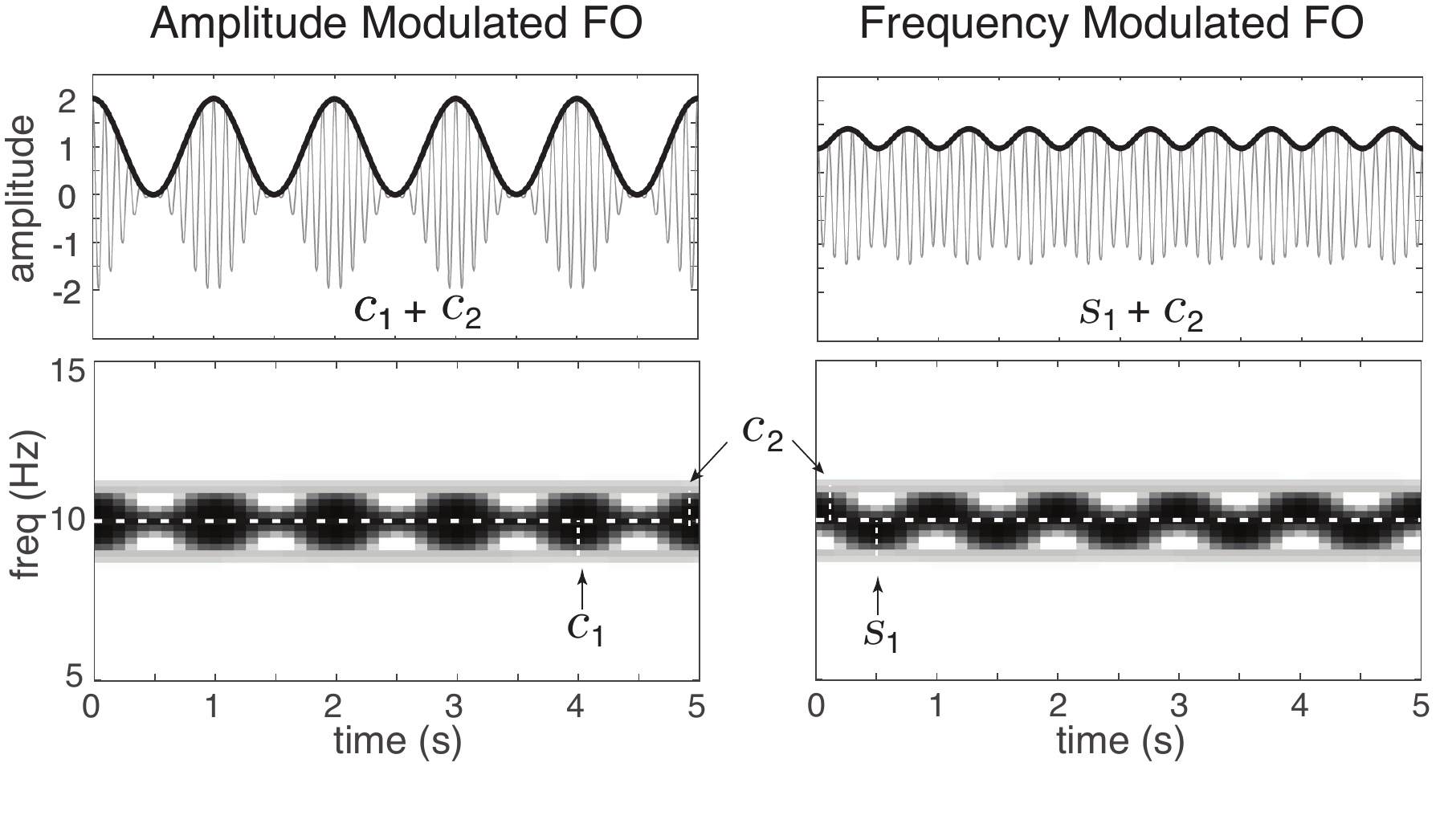}
\caption{
\label{fig:AMFM_illustration}
\add{Amplitude modulation (AM), frequency modulation  (FM) and the smoothness of the bispectrum.  
As described in section \ref{section:FM_bispect}, for an FO occupying a given bandwidth, coupling between the frequency of the FO and the phase of the SO (phase-frequency coupling, PhFC) is reflected in rapidly varying features of the bispectrum in the bandwidth of the FO while phase-amplitude coupling (PAC) is associated with smoothly varying features, illustrated here.
\emph{Left panels:} The modulation of a 10 Hz sinusoid (\emph{gray line}) by a 1 Hz raised cosine \emph{black line} gives rise to two side-band signals, $c_1$ and $c_2$, which correspond to the upper and lower halves of the region highlighted in the time-frequency decomposition (\emph{lower left panel}). 
These can be regarded as separate sinusoidally AM sinusoids, offset by $\pm0.5$ Hz from 10 Hz, whose modulating windows are in phase with each other. 
In the presence of a 1 Hz slow oscillation (SO), the common phase of the respective modulating windows generates two peaks with the same phase \addTwo{(See Fig. \ref{fig:transient_oscillatory}D for an example)}}, modeling a ``smooth'' bispectrum.
\emph{Right panels:} The contrary case, modeling a ``rough'' bispectrum  in which the peaks are at opposite phase,  is illustrated by shifting the phase of the amplitude modulation in the lower band by 90 degrees, substituting $s_1$ for $c_1$ (\emph{lower panel}).
The signal that results from this change (\emph{gray line}) exhibits frequency modulation with comparatively little amplitude modulation (\emph{black line}).
In the presence of the SO, the bispectrum will have the same magnitude as in the previous case, but the phase of the peaks (equivalent to sign in this example) will be opposed. 
}
\end{figure}

\subsection{\add{The Bispectrum and Phase-Frequency Coupling}}
\label{section:FM_bispect}
\add{
So far, our interest in phase-amplitude coupling has led us to focus on smooth bispectra, which gain their smoothness from the windowing of the FO.
As shown in \ref{appendix:PhPC_proof}, conventional  PAC measures aggressively smooth the bispectrum over the FO frequency range, giving them some specificity for bispectral features expected of PAC.}
\add{One might view this as an advantage or disadvantage of PAC measures, depending on context, but it raises the question, what should one make of those sharply-varying features they suppress?}\add{

To gain some insight on the question, we may turn back to the oscillatory model of section \ref{sec:transient_and_oscillatory_models}; in the case of an FO composed of a sinusoidally modulated sinusoid, the bispectrum contained two peaks of the same sign \addTwo{arising from the product of each of the two sidebands in the spectrum of the FO} \delTwo{on either side of the}\addTwo{with the} central ``baseband'' peak \addTwo{(See Fig. \ref{fig:transient_oscillatory}, panel D)}.}
\add{It will be useful to think of the  FO in this case as the sum of two sinusoids offset in frequency from each other by $\theta$ and both modulated by $\cos(\theta/2\:t)$:
\begin{align}
c_1=\cos\left(\theta/2\:t\right)*e^{i(\gamma-\theta/2)t}\quad\mathrm{and}\quad c_2=\cos(\theta/2\:t)*e^{i(\gamma+\theta/2)t}
\end{align}
\begin{align}
\label{EQ_Xam}
f_{\mathrm{AM}}(t)= c_1+c_2=2\cos^2(\theta/2\:t)e^{i\gamma t}
\end{align}}\add{
The resulting FO is illustrated on the left side of Fig. \ref{fig:AMFM_illustration}.

\delTwo{Our model of a sharply varying feature}\addTwo{We may model the opposite of a smooth bispectrum} by inverting \addTwo{one of the two side peaks in the spectrum of the FO within Eq. (\ref{EQ_PureAmpMod})}, which \delTwo{can be done by flipping the sign of one of the side peaks of the FO}\addTwo{has the effect of inverting one of the two peaks in the bispectrum of the FO + SO composite signal:}
\begin{align}
\label{EQ_PureFMod}
\tilde{f}_{\mathrm{FM}}(\omega)= \delta(\omega-\gamma) - \delta(\omega-\gamma-\theta)/2+ \delta(\omega-\gamma+\theta)/2
\end{align} 
Because this change causes the two peaks in the bispectrum to have opposite sign, they will tend to be suppressed by estimators that blur them together. 
The FO in this case is illustrated on the right side of Fig. \ref{fig:AMFM_illustration}. 
As in the former case, this signal may also be written out as the sum of two modulated sinusoids:
\begin{align}
s_1=i\sin\left(\theta/2\:t\right)*e^{i(\gamma-\theta/2)t}\quad\mathrm{and}\quad c_2=\cos(\theta/2\:t)*e^{i(\gamma+\theta/2)t}
\end{align}
The crucial difference lies in the fact that the modulating envelopes of the two components are out of phase with each other, which has two important consequences;
first, the envelope of the composite signal fluctuates over a smaller range than in the previous case:
\begin{align}
\label{EQ_Xfmevn}
\left|X_{\mathrm{FM}}(t)\right|= \left|s_1+c_2\right|=\sqrt{1+\sin^2(\theta t)}
\end{align} 
resulting in amplitude modulation which varies, in this case, over $\left[1,\sqrt{2}\right]$ rather than $[0,2]$.
Second, the instantaneous frequency of the signal gravitates back and forth between between $\gamma+\theta/2$ and $\gamma-\theta/2$ according to which of the two components has the greater amplitude at any moment.
One can therefore attach two interpretations to this case; the first is to regard the components as separate AM signals whose modulating windows happen to be out of phase with each other.  
The second interpretation is that the FO is composed of a single frequency modulated (FM) signal and the bispectrum reflects \emph{phase-frequency coupling} (PhFC) between the frequency of the FO and the phase of the SO. 
Estimators tailored to PhFC might replace the smoothing kernel of PAC and conventional bispectral measures with a high-pass ``sharpening'' kernel.
As with PAC, \delTwo{deciding }whether the interpretation of PhFC \addTwo{is} appropriate in any given case \delTwo{is bound to require consideration of }\addTwo{depends on} other details.\delTwo{, whose}
\delTwo{Further consideration of this question is beyond the scope of the present work, but the exercise shows how bispectral measures might be used to reveal PhFC.} 
}

\section{\add{Methods}}
\subsection{\add{Algorithm}}
\label{sec:algorithm}
\add{
General principles of estimation are reviewed in \ref{section:bispectral_estimators}.
The following section briefly describes the implementation with discretely sampled data used in the present examples. 
The ``direct technique'' of bispectral estimation starts with a windowed Fourier time-frequency decomposition, summing bispectral products over time. 
Although the technique is typically described in its application to the short-time Fourier transform, any time-frequency decomposition, $S[k,n]$, that can be expressed as a bank of single sideband filters might be used.\footnote{\add{Short-time Fourier transform, complex demodulation and single sideband filtering all belong to a family of techniques that are formally equivalent, up to a linear phase term which is negligible for the present purpose because it cancels from the time-invariant bispectral products \citep{kovach2016demodulated}.}} \addTwo{Here $x$ is a discretely sampled signal and $h$ is the time window used by the estimator:} 
\begin{align}
\label{EQ_Sq}
S_q[k,m] = \sum_{n=1}^N{h_{qk}[n]x[n-m]\exp{\left[-i \omega[k]\frac{n}{N}\right]}}
\end{align}
\begin{align}
\label{EQ:bispectral_sum_estimator}
B[j,k] = \sum_{m=0}^{M-1}{S_1[j,m]S_2[k,m]S_3^*[l,m]}
\end{align}
where $j,k$ and $l$ index frequency bands $\omega_1[j],\omega_2[k]$ and $\omega_3[l]$ such that  \[\omega_1[j]+\omega_2[k]=\omega_3[l]\]
or more generally
\[\left|\omega_1[j]+\omega_2[k]-\omega_3[l]\right|<\Delta W_3[l]\]
with $\Delta W_3[l]$ denoting the bandwidth of the $S_3[l,.]$ band.
The use of subindices on $S$ and $h$ here is meant to show that the \addTwo{analysis} windows used \delTwo{in computing}\addTwo{by} the estimator are permitted to vary according to frequency and place in the bispectral product, encompassing the non-standard estimators described in \ref{section:bispectral_estimators}, as well as those with bandwidths that vary over frequency, such as continuous wavelet decompositions \citep{jamsek07}.

 \subsubsection{Normalization}
 Bispectral statistics are shown in normalized form 
 \begin{align}
 \beta[j,k] = \frac{B[j,k]}{A[j,k]}
 \end{align}
 using the \delTwo{alternative }normalization described in \ref{section:bicoherence_alternative}, which takes the sum of bispectral product magnitudes over time as the normalizing term  \citep{hagihira2001practical,kovach2017biased}, given by }\add{
\begin{align}
A[j,k] = \sum_{m=1}^M \left|S_1\left[j,m\right]S_2\left[k,m\right]S_3^*\left[l,m\right]\right|
\end{align}
$A[j,k]$ amounts to the maximum possible value of $|B[j,k]|$, leaving the magnitudes of $S$ unchanged, which is attained when phase remains constant across all terms Eq. (\ref{EQ:bispectral_sum_estimator}).  
 With this normalization, $\beta$ may be interpreted as a weighted-average phase locking value, which avoids some of the ambiguity related to the contribution of amplitudes in more standard forms of bicoherence \citep{srinath14effect,lowet2016quantifying}.}
\add{For a review of a closely-related problem in the context of ordinary coherence, which is directly applicable in this case as well, see \citet{kovach2017biased}.
 }
\subsubsection{\add{Bias Correction}}
\add{
This weighted-average interpretation also allows for a straightforward bias correction \citep{kovach2017biased}, which compensates for amplitude-dependent biases that may vary by frequency \citep{srinath14effect}. 
The bias\delTwo{-}correction subtracts an expected bias from the magnitude bispectrum and renormalizes
\begin{align}
\label{eqn:biascorr}
\left|\beta_{\mathrm{BC}}\right| = \frac{\addTwo{\left|\beta\right|}-\epsilon}{1-\epsilon}
\end{align}
where expected bias is given by }\add{
\begin{align}
\epsilon[j,k] = \frac{\sum_{m=1}^M \left|S_1\left[j,m\right]S_2\left[k,m\right]S_3^*\left[l,m\right]\right|}{\sqrt{\sum_{m=1}^M \left|S_1\left[j,m\right]S_2\left[k,m\right]S_3^*\left[l,m\right]\right|^2}}
\end{align}\addTwo{
\delTwo{For further explanation of the rationale for this, see [ ].}
Note that the use of the magnitude symbol on the left side of Eq. (\ref{eqn:biascorr}) is an abuse of notation as random error might cause  $\left|\beta_\mathrm{BC}\right|$  to become negative. 
The correction can also be applied in a way that preserves phase information:
\begin{align}
\beta_{\mathrm{BC}} = \left|\beta_\mathrm{BC}\right|\frac{\beta}{\left|\beta\right|}
\end{align}
which has the effect of shrinking the positive bias of $\beta$.
For further explanation, see \citet{kovach2017biased}.} 

} 

\subsubsection{\add{Software}}
\add{
In the present examples and simulations, frequency-domain complex demodulation \citep{bingham1967modern} (DBT) was employed as described by \citet{kovach2016demodulated}, using a software implementation for M{\sc atlab} (Mathworks, Nattick, MA) \addTwo{developed by the first author and} available at \url{https://github.com/ckovach/DBT}. 
Polyspectral estimation with \delTwo{the same bandwidths and signal across }\addTwo{fixed bandwidths across all terms for a single signal} is implemented in the function PSPECT, while estimation that accommodates different bandwidths, as well as cross polyspectra, is implemented in PSPECT2.
A DBT implementation of phase-power coupling is given in DBTPAC.

For the demonstrations in figures \ref{fig:275_examp} and \ref{fig:test_signals},  phase-power coupling plots were generated with a bandwidth of 1 Hz for the SO and 40 Hz for the FO.  
Bispectra were computed with a bandwidth of 2 Hz.
}

\subsection{\add{Procedure}}
\add{
Example data shown in Fig. \ref{fig:275_examp} were acquired from an epilepsy patient during a period of clinical monitoring with invasive intracranial recordings.
Data were obtained over 20 minutes during which the patient passively watched a sitcom episode. 
All procedures were approved by the Internal Review Board of the University of Iowa and conducted with voluntary informed consent of the patient. 
For a detailed review of techniques, see \citep{nourski2015invasive}. 
}
\section{\add{Discussion}\color{red}\del{Conclusion}}
\label{sec:discussion}
With respect to the relationship between PAC and the bispectrum, we have shown:
\begin{enumerate}
\item Common measures of PAC based on analytic phase and amplitude are in fact bispectral estimators, and as such provide no unique information beyond what is recovered by standard bispectral estimators. 
\item PAC measures are severely biased with respect to the symmetry of the bispectrum and introduce artificial constraints on the range and resolution of the estimator. 
\end{enumerate}
These limitations provide a clear rationale for favoring standard bispectral estimators in the evaluation of phase-amplitude coupling.
We have further given a more detailed\addTwo{, albeit qualitative,} framework by which to evaluate the presence and nature of PAC from the bispectrum \addTwo{based on the distribution of energy within the bispectrum.}
The signal model developed in pursuit of this framework lends itself to a particularly straightforward interpretation: it describes some recurring feature embedded at times determined by the driving point process.

\subsection{\add{Tuning In to the Bispectrum}}
\label{section:encoding}
\add{Beyond methodological questions, it is worth considering the relevance of this work to theoretical models of the gating and transmission of information in the brain and elsewhere.}\add{
One of the epoch-defining technologies of the last century, radio transmission, developed from the discovery that information can be efficiently conveyed over narrow slices of the radio spectrum.}\add{
In essence, a sender and a receiver agree in advance on the power spectrum of the transmitting signal and tune resonators within their respective devices accordingly.
Information is therefore transmitted within the space of signals of a given power spectrum, and the degrees of freedom within that space, which encode any to-be-transmitted information, correspond to the Fourier-domain phase of the signal.
The crucial problem of overcoming noise is addressed with a ``brute force'' strategy: broadband noise is overwhelmed within a small frequency interval, along with any competing signal, by concentrating the transmission energy in the narrowest practical region of the spectrum, resulting in a tradeoff between signal-to-noise ratio and maximum transmission rate.}

\add{
Starting from the model described in \ref{sec:transient_and_oscillatory_models} and \ref{section:signal_model}, one may carry a similar transmission scheme over to the bispectrum and in return draw forth some promising new features.}\add{ 
The signal space in this scheme is constrained to have a given agreed-upon bispectrum within the $\mathrm{SO \times FO}$ division of the bispectral plane, both with respect to magnitude and phase.}\add{
Transmitted information is encoded by the phase of the $\mathrm{FO}$, which Eq. (\ref{EQ_PPBC4}) allows to vary freely without affecting either the phase- or magnitude-bispectrum.
In contrast, the phase of the $\mathrm{SO}$ and the relative delay of the $\mathrm{FO}$-modulating envelope determine the phase bispectrum within this region, meaning that they are not likewise free to vary.

Three noteworthy features emerge from this scheme:}
\begin{enumerate}
\item \add{The encoding space is specified within a two-dimensional spectral plane, rather than the one dimension afforded by the power spectrum, increasing the number of ways the space may be divided. Among other things, this means that separate channels may transmit information within spectrally overlapping bands without creating interference.}
\item \add{The phase bispectrum preserves information about timing, which allows for the addition of time division to the encoding scheme. 
Eq. (\ref{EQ_PACapprox}) demonstrates how this might work by spacing the modulating window of the $\mathrm{FO}$  at different delays. 
Frequency division is determined by the magnitude bispectrum while time division is determined by the phase bispectrum, thus channels which overlap in both dimensions of the magnitude bispectrum may still avoid interference through a division of the phase bispectrum.}
\item \add{Strategies for overcoming Gaussian noise are not confined to brute-force commandeering of the spectrum in a given region, but may take advantage of the fact that Gaussian processes have vanishing bispectra \cite{Hinich1994}. 
This holds true in the sense of an expectation, meaning in practice, signal-to-noise ratio might be improved through the averaging of repeated transmissions.
}  
\end{enumerate}

\add{
An intuition-friendly summary of this scheme is as follows: it describes the encoding of information within a high-frequency waveform, the $\mathrm{FO}$, which is embedded at a predetermined time relative to a low-frequency carrier waveform the $\mathrm{SO}$; Figure \ref{fig:FreqSmooth} gives an example of how the components of such a signal might appear.}\add{
The $\mathrm{SO}$ may serve as a timing signal which indicates when a new transmission has arrived.
The receiver might identify the transmission by filtering for the $\mathrm{SO}$ and then isolate the $\mathrm{FO}$ with the combination of a second filter and time-window around the  interval encoded by the phase bispectrum.}\add{ 
In this example, the bispectral equivalent of a ``resonator'' is a multistep process involving: (1) identification of the $\mathrm{SO}$ through a suitable filter, which can be derived from the bispectrum as shown by Eq. (\ref{EQ_IRF}).}\add{ 
(2) Peaks in the output of the SO filter triggers the isolation of the $\mathrm{FO}$ through a second filter and time window.}\add{
Finally, if the signal is corrupted by Gaussian noise affecting the output of the $\mathrm{SO}$ filter or the phase of the $\mathrm{FO}$, it may (3) be cleaned up by averaging multiple transmissions, provided the same $\mathrm{FO}$ is also repeated with the $\mathrm{SO}$. }

\add{
Although this scheme raises the complexity of encoding and decoding considerably beyond the simple resonators of ordinary power-spectral transmission, it greatly increases flexibility in the choice of transmitting signal, accommodating transient as well as oscillatory,  spectrally broad as well as narrow signals.}\add{
This scheme also meshes with recent theoretical accounts of the possible functions of PAC. 
Such models emphasize the role of PAC in multiplexing and routing of information through temporal gating and frequency division, both properties that emerge from our consideration of the bispectrum \citep{lisman2005theta,jensen2007cross,young2009coupling,akam2010oscillations,lisman2013theta,akam2014oscillatory,hyafil2015neural}. }\add{By fixing such recent developments within a more formal spectral framework, it may be possible to better elucidate questions of efficiency and optimality as well as to generate predictions about the behavior that efficient and optimal systems should exhibit.}\add{

Common techniques of spectral analysis developed over the first half of the last century in concert with narrowband radio and telegraphic transmission, and it is no accident that they are suited in the first place to the study of spectrally narrow signals.}\add{
Their availability, familiarity and versatility accounts in part for the special attention given to the neural functions of those periodic and oscillatory phenomena about which classical techniques have the most to say \citep{wiener1961cybernetics,buzsaki2006rhythms}, an emphasis carried over into the study of phase-amplitude coupling.}\add{
Yet spectrally broad, often aperiodic phenomena, exemplified most obviously by the action potential and transient evoked responses, also figure prominently in neural signaling at many scales.
The importance of what traditional spectral methods neglect from these signals is becoming increasingly apparent \citep{cole2017brain}.}\add{
The tools afforded by higher-order spectra may go a long way towards illuminating \delTwo{their}\addTwo{the} forms and functions \addTwo{of neural responses }beyond the confines of the narrow passband.
}
\subsection{\add{Future Directions}}
\label{sec:future}
\add{
We have laid out a qualitative framework for understanding and interpreting bispectral statistics.
As one would hope of any useful framework, it brings several interesting questions to the fore, the answering of which must be part of a larger project. 
Our aim has been less to give a comprehensive set of answers than to provoke interest in how the general picture might be further developed. 
No attempt has been made here to define quantitative scores that summarize the qualitative features outlined above. 
The extension of bispectral statistics to the study of phase-frequency coupling was given a cursory and highly schematic treatment and much more might be done to flesh out this application.}\add{

One problem that is particularly ripe for further development is that of distinguishing between bispectra which result from a single recurring feature or the co-occurrence of multiple independent features, and developing a suitable decomposition in the latter case. 
If the underlying processes are independent, the signal bispectrum is a sum of their separate bispectra.}\add{
An important question is how one might go about decomposing the bispectrum or other higher-order spectra according to distinct sets of independent features.
This question is related to one addressed by various blind deconvolution algorithms, a number of which proceed by identifying filters that maximize scalar moments \citep{wiggins1978minimum,donoho1981minimum} (that is, the value of the $k^{th}$ moment at 0 lag); of particular relevance in the case of the bispectrum are those that rely on skewness \citep{paajarvi2004skewness,ovacikli2016recovering}.}
Future work will consider this problem in greater detail.

\section*{Acknowledgements}
\label{Ack}
We wish to thank Phil Gander, Alex Billig and Matthew A. Howard, III.

\bibliographystyle{chicago}
\bibliography{Bibliography}

\begin{thebibliography}{}

\bibitem[\protect\citeauthoryear{Akam and Kullmann}{Akam and
  Kullmann}{2010}]{akam2010oscillations}
Akam, T. and D.~M. Kullmann (2010).
\newblock Oscillations and filtering networks support flexible routing of
  information.
\newblock {\em Neuron\/}~{\em 67\/}(2), 308--320.

\bibitem[\protect\citeauthoryear{Akam and Kullmann}{Akam and
  Kullmann}{2014}]{akam2014oscillatory}
Akam, T. and D.~M. Kullmann (2014).
\newblock Oscillatory multiplexing of population codes for selective
  communication in the mammalian brain.
\newblock {\em Nature Reviews Neuroscience\/}~{\em 15\/}(2), 111--122.

\bibitem[\protect\citeauthoryear{Allen and Rabiner}{Allen and
  Rabiner}{1977}]{allen1977unified}
Allen, J.~B. and L.~R. Rabiner (1977).
\newblock A unified approach to short-time fourier analysis and synthesis.
\newblock {\em Proceedings of the IEEE\/}~{\em 65\/}(11), 1558--1564.

\bibitem[\protect\citeauthoryear{Aru, Aru, Priesemann, Wibral, Lana, Pipa,
  Singer, and Vicente}{Aru et~al.}{2015}]{aru2015untangling}
Aru, J., J.~Aru, V.~Priesemann, M.~Wibral, L.~Lana, G.~Pipa, W.~Singer, and
  R.~Vicente (2015).
\newblock Untangling cross-frequency coupling in neuroscience.
\newblock {\em Current opinion in neurobiology\/}~{\em 31}, 51--61.

\bibitem[\protect\citeauthoryear{Barnett, Johnson, Naitoh, Hicks, and
  Nute}{Barnett et~al.}{1971}]{barnett1971bispectrum}
Barnett, T., L.~Johnson, P.~Naitoh, N.~Hicks, and C.~Nute (1971).
\newblock Bispectrum analysis of electroencephalogram signals during waking and
  sleeping.
\newblock {\em Science\/}~{\em 172\/}(3981), 401--402.

\bibitem[\protect\citeauthoryear{Bartelt, Lohmann, and Wirnitzer}{Bartelt
  et~al.}{1984}]{bartelt1984phase}
Bartelt, H., A.~W. Lohmann, and B.~Wirnitzer (1984).
\newblock Phase and amplitude recovery from bispectra.
\newblock {\em Applied Optics\/}~{\em 23\/}(18), 3121--3129.

\bibitem[\protect\citeauthoryear{Bartlett}{Bartlett}{1963}]{bartlett1963spectral}
Bartlett, M. (1963).
\newblock The spectral analysis of point processes.
\newblock {\em Journal of the Royal Statistical Society. Series B
  (Methodological)\/}, 264--296.

\bibitem[\protect\citeauthoryear{Berman, McDaniel, Liu, Cornew, Gaetz, Roberts,
  and Edgar}{Berman et~al.}{2012}]{berman2012variable}
Berman, J.~I., J.~McDaniel, S.~Liu, L.~Cornew, W.~Gaetz, T.~P. Roberts, and
  J.~C. Edgar (2012).
\newblock Variable bandwidth filtering for improved sensitivity of
  cross-frequency coupling metrics.
\newblock {\em Brain connectivity\/}~{\em 2\/}(3), 155--163.

\bibitem[\protect\citeauthoryear{Bingham, Godfrey, and Tukey}{Bingham
  et~al.}{1967}]{bingham1967modern}
Bingham, C., M.~Godfrey, and J.~Tukey (1967).
\newblock Modern techniques of power spectrum estimation.
\newblock {\em IEEE Transactions on Audio and Electroacoustics\/}~{\em
  15\/}(2), 56--66.

\bibitem[\protect\citeauthoryear{Birkelund, Hanssen, and Powers}{Birkelund
  et~al.}{2003}]{birkelund2003multitaper}
Birkelund, Y., A.~Hanssen, and E.~J. Powers (2003).
\newblock Multitaper estimators of polyspectra.
\newblock {\em Signal Processing\/}~{\em 83\/}(3), 545--559.

\bibitem[\protect\citeauthoryear{Buzsaki}{Buzsaki}{2006}]{buzsaki2006rhythms}
Buzsaki, G. (2006).
\newblock {\em Rhythms of the Brain}.
\newblock Oxford University Press.

\bibitem[\protect\citeauthoryear{Canolty, Edwards, Dalal, Soltani, Nagarajan,
  Kirsch, Berger, Barbaro, and Knight}{Canolty et~al.}{2006}]{canolty2006high}
Canolty, R.~T., E.~Edwards, S.~S. Dalal, M.~Soltani, S.~S. Nagarajan, H.~E.
  Kirsch, M.~S. Berger, N.~M. Barbaro, and R.~T. Knight (2006).
\newblock High gamma power is phase-locked to theta oscillations in human
  neocortex.
\newblock {\em science\/}~{\em 313\/}(5793), 1626--1628.

\bibitem[\protect\citeauthoryear{Chella, Marzetti, Pizzella, Zappasodi, and
  Nolte}{Chella et~al.}{2014}]{chella2014third}
Chella, F., L.~Marzetti, V.~Pizzella, F.~Zappasodi, and G.~Nolte (2014).
\newblock Third order spectral analysis robust to mixing artifacts for mapping
  cross-frequency interactions in eeg/meg.
\newblock {\em Neuroimage\/}~{\em 91}, 146--161.

\bibitem[\protect\citeauthoryear{Cohen}{Cohen}{1989}]{Cohen1989}
Cohen, L. (1989, Jul).
\newblock Time-frequency distributions-a review.
\newblock {\em Proceedings of the IEEE\/}~{\em 77\/}(7), 941--981.

\bibitem[\protect\citeauthoryear{Cole and Voytek}{Cole and
  Voytek}{2017}]{cole2017brain}
Cole, S.~R. and B.~Voytek (2017).
\newblock Brain oscillations and the importance of waveform shape.
\newblock {\em Trends in Cognitive Sciences\/}~{\em 21\/}(2), 137 -- 149.

\bibitem[\protect\citeauthoryear{Daley and Vere-Jones}{Daley and
  Vere-Jones}{2003}]{daley2003introductionI}
Daley, D.~J. and D.~Vere-Jones (2003).
\newblock {\em An Introduction to the Theory of Point Processes. Volume I:
  Elementary Theory and Methods\/} (2nd ed.).
\newblock Springer Science \& Business Media.

\bibitem[\protect\citeauthoryear{Donoho}{Donoho}{1981}]{donoho1981minimum}
Donoho, D. (1981).
\newblock On minimum entropy deconvolution.
\newblock {\em Applied Time Series Analysis 2\/}, 561--608.

\bibitem[\protect\citeauthoryear{Dumermuth, Huber, Kleiner, and
  Gasser}{Dumermuth et~al.}{1971}]{dumermuth1971analysis}
Dumermuth, G., P.~Huber, B.~Kleiner, and T.~Gasser (1971).
\newblock Analysis of the interrelations between frequency bands of the eeg by
  means of the bispectrum a preliminary study.
\newblock {\em Electroencephalography and clinical neurophysiology\/}~{\em
  31\/}(2), 137--148.

\bibitem[\protect\citeauthoryear{Fackrell, White, Hammond, Pinnington, and
  Parsons}{Fackrell et~al.}{1995}]{fackrell1995interpretation}
Fackrell, J., P.~White, J.~Hammond, R.~Pinnington, and A.~Parsons (1995).
\newblock The interpretation of the bispectra of vibration signalsÑ: I. theory.
\newblock {\em Mechanical Systems and Signal Processing\/}~{\em 9\/}(3),
  257--266.

\bibitem[\protect\citeauthoryear{Fonoliosa and Nikias}{Fonoliosa and
  Nikias}{1993}]{fonoliosa1993wigner}
Fonoliosa, J. and C.~Nikias (1993).
\newblock Wigner higher order moment spectra: definition, properties,
  computation and application to transient signal analysis.
\newblock {\em IEEE Transactions on Signal Processing\/}~{\em 41\/}(1), 245.

\bibitem[\protect\citeauthoryear{Gan, Glass, Windsor, Payne, Rosow, Sebel, and
  Manberg}{Gan et~al.}{1997}]{gan1997bispectral}
Gan, T.~J., P.~S. Glass, A.~Windsor, F.~Payne, C.~Rosow, P.~Sebel, and
  P.~Manberg (1997).
\newblock Bispectral index monitoring allows faster emergence and improved
  recovery from propofol, alfentanil, and nitrous oxide anesthesia.
\newblock {\em The Journal of the American Society of Anesthesiologists\/}~{\em
  87\/}(4), 808--815.

\bibitem[\protect\citeauthoryear{Gerr}{Gerr}{1988}]{gerr88}
Gerr, N.~L. (1988, Mar).
\newblock Introducing a third-order wigner distribution.
\newblock {\em Proceedings of the IEEE\/}~{\em 76\/}(3), 290--292.

\bibitem[\protect\citeauthoryear{Godfrey}{Godfrey}{1965}]{Godfrey65}
Godfrey, M.~D. (1965).
\newblock An exploratory study of the bi-spectrum of economic time series.
\newblock {\em Journal of the Royal Statistical Society. Series C (Applied
  Statistics)\/}~{\em 14\/}(1), 48--69.

\bibitem[\protect\citeauthoryear{Hagihira, Takashina, Mori, Mashimo, and
  Yoshiya}{Hagihira et~al.}{2001}]{hagihira2001practical}
Hagihira, S., M.~Takashina, T.~Mori, T.~Mashimo, and I.~Yoshiya (2001).
\newblock Practical issues in bispectral analysis of electroencephalographic
  signals.
\newblock {\em Anesthesia \& Analgesia\/}~{\em 93\/}(4), 966--970.

\bibitem[\protect\citeauthoryear{Hanssen and Scharf}{Hanssen and
  Scharf}{2003}]{hanssen2003theory}
Hanssen, A. and L.~L. Scharf (2003).
\newblock A theory of polyspectra for nonstationary stochastic processes.
\newblock {\em IEEE Transactions on Signal Processing\/}~{\em 51\/}(5),
  1243--1252.

\bibitem[\protect\citeauthoryear{Hinich}{Hinich}{1994}]{Hinich1994}
Hinich, M.~J. (1994, Dec).
\newblock Higher order cumulants and cumulant spectra.
\newblock {\em Circuits, Systems and Signal Processing\/}~{\em 13\/}(4),
  391--402.

\bibitem[\protect\citeauthoryear{Hlawatsch and Boudreaux-Bartels}{Hlawatsch and
  Boudreaux-Bartels}{1992}]{hlawatsch1992linear}
Hlawatsch, F. and G.~F. Boudreaux-Bartels (1992).
\newblock Linear and quadratic time-frequency signal representations.
\newblock {\em IEEE Signal Processing Magazine\/}~{\em 9\/}(2), 21--67.

\bibitem[\protect\citeauthoryear{Hyafil}{Hyafil}{2015}]{hyafil2015misidentifications}
Hyafil, A. (2015).
\newblock Misidentifications of specific forms of cross-frequency coupling:
  three warnings.
\newblock {\em Frontiers in neuroscience\/}~{\em 9}.

\bibitem[\protect\citeauthoryear{Hyafil, Giraud, Fontolan, and Gutkin}{Hyafil
  et~al.}{2015}]{hyafil2015neural}
Hyafil, A., A.-L. Giraud, L.~Fontolan, and B.~Gutkin (2015).
\newblock Neural cross-frequency coupling: connecting architectures,
  mechanisms, and functions.
\newblock {\em Trends in neurosciences\/}~{\em 38\/}(11), 725--740.

\bibitem[\protect\citeauthoryear{Jam\ifmmode~\check{s}\else \v{s}\fi{}ek,
  Stefanovska, and McClintock}{Jam\ifmmode~\check{s}\else \v{s}\fi{}ek
  et~al.}{2007}]{jamsek07}
Jam\ifmmode~\check{s}\else \v{s}\fi{}ek, J., A.~Stefanovska, and P.~V.~E.
  McClintock (2007, Oct).
\newblock Wavelet bispectral analysis for the study of interactions among
  oscillators whose basic frequencies are significantly time variable.
\newblock {\em Phys. Rev. E\/}~{\em 76}, 046221.

\bibitem[\protect\citeauthoryear{Jensen and Colgin}{Jensen and
  Colgin}{2007}]{jensen2007cross}
Jensen, O. and L.~L. Colgin (2007).
\newblock Cross-frequency coupling between neuronal oscillations.
\newblock {\em Trends in cognitive sciences\/}~{\em 11\/}(7), 267--269.

\bibitem[\protect\citeauthoryear{Jirsa and M{\"u}ller}{Jirsa and
  M{\"u}ller}{2013}]{jirsa2013}
Jirsa, V. and V.~M{\"u}ller (2013).
\newblock Cross-frequency coupling in real and virtual brain networks.
\newblock {\em Frontiers in Computational Neuroscience\/}~{\em 7}, 78.

\bibitem[\protect\citeauthoryear{Kearse, Rosow, Zaslavsky, Connors, Dershwitz,
  and Denman}{Kearse et~al.}{1998}]{kearse1998bispectral}
Kearse, L.~A., C.~Rosow, A.~Zaslavsky, P.~Connors, M.~Dershwitz, and W.~Denman
  (1998).
\newblock Bispectral analysis of the electroencephalogram predicts conscious
  processing of information during propofol sedation and hypnosis.
\newblock {\em The Journal of the American Society of Anesthesiologists\/}~{\em
  88\/}(1), 25--34.

\bibitem[\protect\citeauthoryear{Kim and Powers}{Kim and
  Powers}{1979}]{kim1979}
Kim, Y.~C. and E.~J. Powers (1979, June).
\newblock Digital bispectral analysis and its applications to nonlinear wave
  interactions.
\newblock {\em IEEE Transactions on Plasma Science\/}~{\em 7\/}(2), 120--131.

\bibitem[\protect\citeauthoryear{Kovach}{Kovach}{2017}]{kovach2017biased}
Kovach, C.~K. (2017).
\newblock A biased look at phase locking: Brief critical review and proposed
  remedy.
\newblock {\em IEEE Transactions on Signal Processing\/}~{\em 65\/}(17),
  4468--4480.

\bibitem[\protect\citeauthoryear{Kovach and Gander}{Kovach and
  Gander}{2016}]{kovach2016demodulated}
Kovach, C.~K. and P.~E. Gander (2016).
\newblock The demodulated band transform.
\newblock {\em Journal of neuroscience methods\/}~{\em 261}, 135--154.

\bibitem[\protect\citeauthoryear{Kramer, Tort, and Kopell}{Kramer
  et~al.}{2008}]{kramer2008sharp}
Kramer, M.~A., A.~B. Tort, and N.~J. Kopell (2008).
\newblock Sharp edge artifacts and spurious coupling in eeg frequency
  comodulation measures.
\newblock {\em Journal of neuroscience methods\/}~{\em 170\/}(2), 352--357.

\bibitem[\protect\citeauthoryear{Lisman}{Lisman}{2005}]{lisman2005theta}
Lisman, J. (2005).
\newblock The theta/gamma discrete phase code occuring during the hippocampal
  phase precession may be a more general brain coding scheme.
\newblock {\em Hippocampus\/}~{\em 15\/}(7), 913--922.

\bibitem[\protect\citeauthoryear{Lisman and Jensen}{Lisman and
  Jensen}{2013}]{lisman2013theta}
Lisman, J.~E. and O.~Jensen (2013).
\newblock The theta-gamma neural code.
\newblock {\em Neuron\/}~{\em 77\/}(6), 1002--1016.

\bibitem[\protect\citeauthoryear{Lowet, Roberts, Bonizzi, Karel, and
  De~Weerd}{Lowet et~al.}{2016}]{lowet2016quantifying}
Lowet, E., M.~J. Roberts, P.~Bonizzi, J.~Karel, and P.~De~Weerd (2016).
\newblock Quantifying neural oscillatory synchronization: A comparison between
  spectral coherence and phase-locking value approaches.
\newblock {\em PloS one\/}~{\em 11\/}(1).

\bibitem[\protect\citeauthoryear{Lozano-Soldevilla, ter Huurne, and
  Oostenveld}{Lozano-Soldevilla et~al.}{2016}]{lozano2016neuronal}
Lozano-Soldevilla, D., N.~ter Huurne, and R.~Oostenveld (2016).
\newblock Neuronal oscillations with non-sinusoidal morphology produce spurious
  phase-to-amplitude coupling and directionality.
\newblock {\em Frontiers in computational neuroscience\/}~{\em 10}, 87.

\bibitem[\protect\citeauthoryear{Myles, Leslie, McNeil, Forbes, Chan, Group,
  et~al.}{Myles et~al.}{2004}]{myles2004bispectral}
Myles, P., K.~Leslie, J.~McNeil, A.~Forbes, M.~Chan, B.-A.~T. Group, et~al.
  (2004).
\newblock Bispectral index monitoring to prevent awareness during anaesthesia:
  the b-aware randomised controlled trial.
\newblock {\em The lancet\/}~{\em 363\/}(9423), 1757--1763.

\bibitem[\protect\citeauthoryear{Nikias and Mendel}{Nikias and
  Mendel}{1993}]{nikias1993signal}
Nikias, C.~L. and J.~M. Mendel (1993).
\newblock Signal processing with higher-order spectra.
\newblock {\em IEEE signal processing magazine\/}~{\em 10\/}(3), 10--37.

\bibitem[\protect\citeauthoryear{Nikias and Raghuveer}{Nikias and
  Raghuveer}{1987}]{nikias1987bispectrum}
Nikias, C.~L. and M.~R. Raghuveer (1987).
\newblock Bispectrum estimation: A digital signal processing framework.
\newblock {\em Proceedings of the IEEE\/}~{\em 75\/}(7), 869--891.

\bibitem[\protect\citeauthoryear{Nourski and Howard~III}{Nourski and
  Howard~III}{2015}]{nourski2015invasive}
Nourski, K.~V. and M.~A. Howard~III (2015).
\newblock Invasive recordings in the human auditory cortex.
\newblock {\em Handb. Clin. Neurol\/}~{\em 129}, 225--244.

\bibitem[\protect\citeauthoryear{Ovac{\i}kl{\i}, P{\"a}{\"a}j{\"a}rvi, LeBlanc,
  and Carlson}{Ovac{\i}kl{\i} et~al.}{2016}]{ovacikli2016recovering}
Ovac{\i}kl{\i}, A.~K., P.~P{\"a}{\"a}j{\"a}rvi, J.~P. LeBlanc, and J.~E.
  Carlson (2016, March).
\newblock Recovering periodic impulsive signals through skewness maximization.
\newblock {\em IEEE Transactions on Signal Processing\/}~{\em 64\/}(6),
  1586--1596.

\bibitem[\protect\citeauthoryear{P{\"a}{\"a}j{\"a}rvi and
  Leblanc}{P{\"a}{\"a}j{\"a}rvi and Leblanc}{2004}]{paajarvi2004skewness}
P{\"a}{\"a}j{\"a}rvi, P. and J.~Leblanc (2004).
\newblock Skewness maximization for impulsive sources in blind deconvolution.
\newblock In {\em Nordic Signal Processing Symposium: 09/06/2004-11/06/2004},
  pp.\  304--307. Helsinki University of Technology.

\bibitem[\protect\citeauthoryear{Sasaki, Sato, and Yamashita}{Sasaki
  et~al.}{1975}]{sasaki1975minimum}
Sasaki, K., T.~Sato, and Y.~Yamashita (1975).
\newblock Minimum bias windows for bispectral estimation.
\newblock {\em Journal of Sound and Vibration\/}~{\em 40\/}(1), 139--148.

\bibitem[\protect\citeauthoryear{Schack, Vath, Petsche, Geissler, and
  M{\"o}ller}{Schack et~al.}{2002}]{schack2002phase}
Schack, B., N.~Vath, H.~Petsche, H.-G. Geissler, and E.~M{\"o}ller (2002).
\newblock Phase-coupling of theta--gamma eeg rhythms during short-term memory
  processing.
\newblock {\em International Journal of Psychophysiology\/}~{\em 44\/}(2),
  143--163.

\bibitem[\protect\citeauthoryear{Scheffer-Teixeira and Tort}{Scheffer-Teixeira
  and Tort}{2016}]{Sheffer-Teixera16}
Scheffer-Teixeira, R. and A.~B. Tort (2016, dec).
\newblock On cross-frequency phase-phase coupling between theta and gamma
  oscillations in the hippocampus.
\newblock {\em eLife\/}~{\em 5}, e20515.

\bibitem[\protect\citeauthoryear{Sheremet, Burke, and Maurer}{Sheremet
  et~al.}{2016}]{Sheremet4218}
Sheremet, A., S.~N. Burke, and A.~P. Maurer (2016).
\newblock Movement enhances the nonlinearity of hippocampal theta.
\newblock {\em Journal of Neuroscience\/}~{\em 36\/}(15), 4218--4230.

\bibitem[\protect\citeauthoryear{Shils, Litt, Skolnick, and Stecker}{Shils
  et~al.}{1996}]{shils1996bispectral}
Shils, J., M.~Litt, B.~Skolnick, and M.~Stecker (1996).
\newblock Bispectral analysis of visual interactions in humans.
\newblock {\em Electroencephalography and clinical neurophysiology\/}~{\em
  98\/}(2), 113--125.

\bibitem[\protect\citeauthoryear{Sigl and Chamoun}{Sigl and
  Chamoun}{1994}]{sigl1994introduction}
Sigl, J.~C. and N.~G. Chamoun (1994).
\newblock An introduction to bispectral analysis for the electroencephalogram.
\newblock {\em Journal of clinical monitoring\/}~{\em 10\/}(6), 392--404.

\bibitem[\protect\citeauthoryear{Srinath and Ray}{Srinath and
  Ray}{2014}]{srinath14effect}
Srinath, R. and S.~Ray (2014).
\newblock Effect of amplitude correlations on coherence in the local field
  potential.
\newblock {\em Journal of Neurophysiology\/}~{\em 112\/}(4), 741--751.

\bibitem[\protect\citeauthoryear{Swami}{Swami}{1991}]{swami88}
Swami, A. (1991, Apr).
\newblock Third-order wigner distributions: definitions and properties.
\newblock In {\em [Proceedings] ICASSP 91: 1991 International Conference on
  Acoustics, Speech, and Signal Processing}, pp.\  3081--3084 vol.5.

\bibitem[\protect\citeauthoryear{Tort, Komorowski, Eichenbaum, and Kopell}{Tort
  et~al.}{2010}]{Tort2010}
Tort, A. B.~L., R.~Komorowski, H.~Eichenbaum, and N.~Kopell (2010).
\newblock Measuring phase-amplitude coupling between neuronal oscillations of
  different frequencies.
\newblock {\em Journal of Neurophysiology\/}~{\em 104\/}(2), 1195--1210.

\bibitem[\protect\citeauthoryear{van Driel, Cox, and Cohen}{van Driel
  et~al.}{2015}]{van2015phase}
van Driel, J., R.~Cox, and M.~X. Cohen (2015).
\newblock Phase-clustering bias in phase--amplitude cross-frequency coupling
  and its removal.
\newblock {\em Journal of neuroscience methods\/}~{\em 254}, 60--72.

\bibitem[\protect\citeauthoryear{Vaz, Yaffe, Wittig, Inati, and Zaghloul}{Vaz
  et~al.}{2017}]{vaz2017dual}
Vaz, A.~P., R.~B. Yaffe, J.~H. Wittig, S.~K. Inati, and K.~A. Zaghloul (2017).
\newblock Dual origins of measured phase-amplitude coupling reveal distinct
  neural mechanisms underlying episodic memory in the human cortex.
\newblock {\em Neuroimage\/}~{\em 148}, 148--159.

\bibitem[\protect\citeauthoryear{Wiener}{Wiener}{1961}]{wiener1961cybernetics}
Wiener, N. (1961).
\newblock {\em Cybernetics or Control and Communication in the Animal and the
  Machine}, Volume~25.
\newblock MIT press.

\bibitem[\protect\citeauthoryear{Wiggins}{Wiggins}{1978}]{wiggins1978minimum}
Wiggins, R.~A. (1978).
\newblock Minimum entropy deconvolution.
\newblock {\em Geoexploration\/}~{\em 16\/}(1-2), 21--35.

\bibitem[\protect\citeauthoryear{Young and Eggermont}{Young and
  Eggermont}{2009}]{young2009coupling}
Young, C.~K. and J.~J. Eggermont (2009).
\newblock Coupling of mesoscopic brain oscillations: recent advances in
  analytical and theoretical perspectives.
\newblock {\em Progress in neurobiology\/}~{\em 89\/}(1), 61--78.

\end{thebibliography}


\begin{appendix}

\section{Properties of the Bispectrum}
\label{section:bispectral_properties}
Here we review a handful of properties that will be called upon in developing the transient model of section \ref{sec:transient_and_oscillatory_models} and \ref{section:signal_model}. \add{Unless otherwise specified, variables $r,s,t$ refer to time and $\tau$ to time lags.When $\delta$ is within the argument of a function it is also a lag variable, while outside of an argument it will represent the Dirac delta function. Frequency variables will be represented with $\omega,\nu,\eta,\xi$ and $\lambda$.}

\subsection{Symmetry}
\label{SEC_symmetry}
The third moment is unchanged under a permutation of the time delays, $\tau_1$ and $\tau_2$, over the three instances of $X$ in Eq. (\ref{EQ_ambi3rd}), creating a six-fold symmetry, illustrated in Figure \ref{fig:symmetry},  with the following relations: 
\begin{align}
\label{EQ_symmetry}
\mu_3(\tau_1,\tau_2) = \mu_3(\tau_2,\tau_1) = \mu_3(-\tau_1,\tau_2-\tau_1)
\end{align}
Because the spectrum of a real-valued signal is conjugate symmetric about the origin, the 6-fold symmetry of the third moment becomes a 12-fold conjugate symmetry in the bispectrum under the following relations:
\begin{align}
\label{EQ_Fsymmetry}
\tilde{\mu}_3(\omega_1,\omega_2) = \tilde{\mu}_3(\omega_2,\omega_1) = \tilde{\mu}_3(-\omega_1-\omega_2,\omega_2)= \tilde{\mu}^*_3(-\omega_1,-\omega_2)
\end{align}
For a real signal, the full bispectrum may be recovered from estimates within any one of the symmetry regions. These relations apply as well to the frequency dimensions of the third-order Wigner-Ville distribution, $W_3$.

\subsection{Convolution  }
\label{sec:convolution}
The bispectrum inherits a number of basic spectral properties, among which are those related to convolution. 
Because the convolution of two signals in time equates to multiplication in the spectral domain, it is directly apparent that the bispectrum of a signal formed from the convolution of two other signals is likewise the product of their separate bispectra, provided the signals are third-order independent or deterministic; that is, 
\begin{align}
\label{EQ_convprop}
B[X\circ Y] = B[X]B[Y]
\end{align}
for third-order independent $X,Y$. 

\subsection{Multiplication}

The reverse relationship, by which the product of two independent signals in time becomes a convolution in the frequency domain holds generally for the bispectrum only if at least one of the signals is also third-order stationary; that is
\begin{align}
\label{EQ_convprop2}
B[XY] = B[X]\circ B[Y]
\end{align}
if $X$ and $Y$ are third-order independent and if $X$ or $Y$ is third-order stationary.
Because of the stationarity requirement, this relation applies only in trivial cases (i.e. $X(t) = c$) when both signals are deterministic\addTwo{, such that $E\left[X(t)\right]=X(t)$.} 

The relationship does, however, hold generally for the bispectral Wigner-Ville distribution; that is 
\begin{align}
\label{EQ_WVconvprop}
W_3[XY] = W_3[X]\circ_\omega W_3[Y]
\end{align}
for third-order independent $X$ and $Y$, where $\circ_\omega$ denotes convolution over the frequency dimension.
Then 
\begin{align}
\label{EQ_WVconvprop2}
B[XY] = \mathrm{E}\left[\int{W_3[X]\circ_\omega W_3[Y]\mathop{dt}}\right]
\end{align}
For third-order stationary $X$, time drops out of the expectation as $ \mathrm{E}\left[W_3[X]\right]=\tilde{\mu}_3[X]$, so that 
\begin{align}
\label{EQ_WVconvprop_stationary}
B[XY] &= \mathrm{E}\left[\int{W_3[X]\circ_\omega W_3[Y]\mathop{dt}}\right]\\
	&= \tilde{\mu}_3[X]\circ\int{ \mathrm{E}\left[ W_3[Y]\right]\mathop{dt}} = \tilde{\mu}_3 [X]\circ B[Y]
\end{align}
if $X$ is stationary and third-order independent of $Y$.

\subsection{Alternative Forms}
The lags in Eq. (\ref{EQ_ambi3rd}) can be defined in different ways while preserving the essential properties of $W_3$ \citep{swami88}; it will at times be useful to apply a change of variables giving:
\begin{align}
\label{EQ_wignerville}
	 W_3(\nu_1,\nu_2,t) = 
	 		\iint{ X(t+\delta)X(t-\delta)X(t-\tau)e^{-i\nu_1\delta-i\nu_2\tau}\mathop{d\tau}\mathop{d\delta}
						} 
\end{align}
where $\nu_1\equiv\omega_1-\omega_2$ and $\nu_2\equiv\omega_1+\omega_2$, $\delta\equiv(\tau_1-\tau_2)/2$, and $\tau\equiv(\tau_1+\tau_2)/2$. Where necessary, the distinction between these forms will be made through a similar abuse of notation in the arguments.

\begin{figure} 
\centering
\includegraphics[width=75mm]{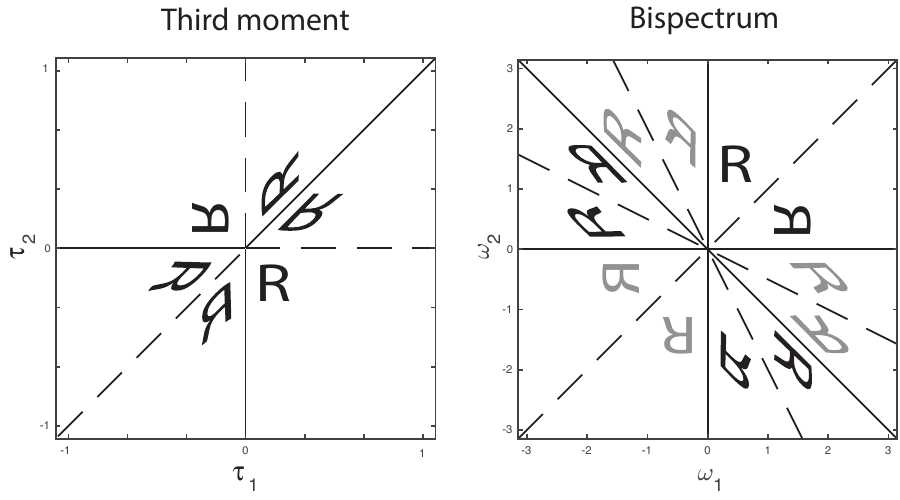}
\caption{\label{fig:symmetry} 
Symmetry of the third moment (\emph{left panel}) and the bispectrum (\emph{right panel}). \add{
For a real-valued signal, the six-fold symmetry of the third moment becomes a twelve-fold symmetry in which 6 pairs of regions are conjugate symmetric (complex conjugate indicated in gray).}
}
\end{figure}

Similarly, in the spectral domain, it will later become useful to consider  the change of variables $\omega^\prime_1 = \omega_1$ and $\omega^\prime_2 = \omega_{\add{2}}+\omega_{\add{1}}/2$. This change effectively modifies the definition of the bispectrum to a form that is symmetrical around $\omega_2$ :
\begin{align}
\label{EQ_bispectsym}
	 B^\prime(\omega^\prime_1,\omega^\prime_2) =  \mathrm{E}\left[ \tilde{X}(\omega^\prime_1)\tilde{X}(\omega^\prime_2-\omega^\prime_1/2)\tilde{X}^*(\omega^\prime_2+\omega^\prime_1/2)\right]
\end{align}\add{
Under this change of variables, the bispectrum assumes a radial symmetry with each of the axes shown in Fig. \ref{fig:symmetry} separated by 30 degrees.} 
The corresponding change in the time-domain involves a modification of the lags in the 3rd moment, $\tau^\prime_1=\tau_1\add{-\tau_2/2}$ and $\tau^\prime_2=\add{\tau_2}$, adding $t^\prime=t+\add{\tau_2/2}$ for the sake of symmetry gives
\begin{align}
\label{EQ_musym}
\mu^\prime_3(\tau^\prime_1,\tau^\prime_2,t^\prime)=E\left[X(t^\prime+\add{\tau^\prime_1})X(t^\prime+\add{\frac{\tau^\prime_2}{2}})X(t^\prime-\add{\frac{\tau^\prime_2}{2}})\right] 
\end{align}

\section{Bispectral Estimators}
\label{section:bispectral_estimators}
A central concern here is how to obtain an estimator of $B$ for a third-order stationary signal. The extension to non-stationary estimation parallels that of ordinary spectral estimation and will not be considered separately here. The most straightforward approaches follow the example set by conventional methods for computing Fourier power spectra \citep{nikias1987bispectrum,birkelund2003multitaper}. The so-called ``indirect'' method first estimates the third moment of the signal in the time domain and  windows it, giving $\hat{\mu}^H_3=\hat{\mu}_3(\tau_1,\tau_2)H(\tau_1,\tau_2)$, from which the smoothed bispectrum is obtained with a two-dimensional Fourier transform. The ``direct'' method first computes a series of windowed Fourier spectra, $\chi_j(\omega)$, and averages over the product $\chi_j(\omega_1)\chi_j(\omega_2)\chi_j^*(\omega_1+\omega_2)$. 
Both approaches generate a largely equivalent family of estimators. 

To minimize bias in the estimate and give an estimate with the proper symmetry, it is usually required that $H$ itself exhibit the same symmetry as the third moment, given in Eq. (\ref{EQ_symmetry}). Windows should also be real-valued in time and frequency and non-negative in the frequency domain \citep{sasaki1975minimum}.  
Two-dimensional windows fulfilling these requirements may be constructed in either the time or frequency domains from a window defined in one dimension, $h(t)$: 
\begin{align}
\label{EQ_symmetry2}
H(\tau_1,\tau_2) = h(\tau_1)h(\tau_2)h(\tau_1-\tau_2)
\qquad \mathrm{ or }\qquad
\tilde{H}(\omega_1,\omega_2)=\tilde{h}(\omega_1)\tilde{h}(\omega_2)\tilde{h}(\omega_1+\omega_2)
\end{align}
Although specific windows are shown to be optimal for minimizing bias of bispectral estimators \citep{sasaki1975minimum}, we will consider a more general case, which also includes windows for which symmetry relation may not hold.
We will relate PAC measures to bispectral estimators that lack the proper symmetry constraint, which motivates dropping the constraint in the present discussion.
It will still be assumed that the windows are non-negative and real-valued in the frequency domain and symmetric under a 180 degree rotation,  although the complex conjugate will be indicated where appropriate for notational consistency. 

The ``direct'' estimator most commonly begins with a series of windowed Fourier transforms, evenly spaced in time (here treated as continuous for simplicity), which may be regarded equivalently as the outcome of a bank of filters whose outputs are analytic signals shifted in frequency (i.e. complex demodulates) obtained by bandpass-filtering the original signal \citep{Godfrey65, bingham1967modern,kovach2016demodulated}:
\begin{align}
\begin{split}
\label{EQ_tf}
	 \chi_h(\omega,t) &=  \int{h(s-t)X(s)e^{-i\omega s}\mathop{ds}
			}
\end{split}
\end{align}
An estimate of the bispectrum is obtained from this time-frequency representation as
\begin{align}
\begin{split}
\label{EQ_bispectest0}
	 \hat{B}&(\omega_1,\omega_2) = \\ & \frac{1}{T}\int_T{\chi_h(\omega_1,s)\chi_h(\omega_2,s)\chi_h^*(\omega_1+\omega_2,s)\mathop{ds}
			}\\
			= & \frac{1}{T}\iiint{X(t+\tau_1)X(t+\tau_2)X(t)e^{-i\omega_1\tau_1-i\omega_2\tau_2}\int{h(s+\tau_1)h(s+\tau_2)h(s)\mathop{ds}}\mathop{d\tau_1}\mathop{d\tau_2}\mathop{dt}
			}
\end{split}
\end{align}
The last integrand \add{in (\ref{EQ_bispectest0})} defines the smoothing window:
\begin{align}
\begin{split}
\label{EQ_bcsmooth}
	H(\tau_1,\tau_2) =  & \int{h(s+\tau_1)h(s+\tau_2)h(s)\mathop{ds}
			}\\
	     =  & \int{\tilde{h}(\xi_1)\tilde{h}(\xi_2)\tilde{h}^*(\xi_1+\xi_2)e^{i\xi_1\tau_1+i\xi_2\tau_2}\mathop{d\xi_1}\mathop{d\xi_2}		}\\
\end{split}
\end{align}
Comparing this to Eq. (\ref{EQ_symmetry2}) makes it clear that the resulting estimator fulfills the symmetry requirement of Eq. (\ref{EQ_symmetry}).
Combining these results:
\begin{align}
\begin{split}
\label{EQ_bispectest}
	 B&(\omega_1,\omega_2) =  \iiint{W(\xi_1,\xi_2,t)\tilde{H}(\xi_1-\omega_1,\xi_2-\omega_2)\mathop{d\xi_1}\mathop{d\xi_2}\mathop{dt}
			}
\end{split}
\end{align}
which makes the smoothing property of the bispectral estimator explicit.
With a time-centering change of variables, this relation also gives a time-varying estimate of the bispectrum as a smoothing of the third-order Wigner-Ville distribution, absent the integration over time  \citep{fonoliosa1993wigner,hanssen2003theory}. As in the second-order case, suppressing cross terms in the time-varying spectrum requires some additional smoothing over time.

\subsection{Bicoherence}
 Each of the estimators considered above can be regarded as an inner product between two signals, the first being the higher analytic band, $\chi_3(\omega_1+\omega_2,t)$ and the second obtained by multiplying the two lower bands $\chi_1(\omega_1,t)\chi_2(\omega_2,t)$. 
The degree of alignment between these two terms is often used as a normalized measure of dependence, which is obtained in the same way as for an ordinary correlation, by dividing the inner product with the magnitudes of the separate terms \cite{kim1979}:
\begin{align}
\label{EQ_Bicoh_def1}
\beta(\omega_1,\omega_2)= \frac{\hat{B}(\omega_1,\omega_2)}{
				\sqrt{
					\int{
					    \left|\chi_1(\omega_1,t)\chi_2(\omega_2,t)\right|^2
					    \mathop{dt} }}
					   \sqrt{\int{	
					    \left|\chi_3(\omega_1+\omega_2,s)\right|^2
					    \mathop{ds} }
				}
			}
\end{align}
This is a commonly used definition of \emph{bicoherence}. 

\subsubsection{Alternative Definition}
\label{section:bicoherence_alternative}
It is worth noting that this common definition of  bicoherence does not share the full set of symmetry properties with the bispectrum because the normalizing term is not likewise symmetric. 
Whereas $\beta(\omega_1,\omega_2)=\beta(\omega_2,\omega_1)$, it is not generally true that $\beta(\omega_1,\omega_2)=\beta(-\omega_1-\omega_2,\omega_2)$. Some alternative definitions of bicoherence do preserve the appropriate symmetry. 
The following defines an alternative index \citep{hagihira2001practical}:
\begin{align}
\label{EQ_Bicoh_def2}
\beta_\phi(\omega_1,\omega_2)= \frac{\hat{B}(\omega_1,\omega_2)}{				
					\int{					
					   \left|\chi_1(\omega_1,t)
					   \chi_2(\omega_2,t)			    
					    \chi^*_3(\omega_1+\omega_2,t)\right|
					    }\mathop{dt} 
			}
\end{align}
This index may be interpreted as a weighted-average vector strength over the phase component of the bispectrum, thus it gives a measure of phase locking \citep{kovach2017biased}.

\subsection{Estimators with Different Analysis Filters}
\label{appendix:DifferentFilters}
Eq. (\ref{EQ_bispectest0}) constrains the analysis window for each of the three bands used in the estimate to be identical, which had the benefit of yielding an estimator with the proper symmetry. 
While the  symmetry conditions in (\ref{EQ_symmetry}) are usually regarded as a criterion of admissibility for windows used in estimating the bispectrum, there is in fact no essential reason why one may not smooth the bispectrum with windows that do not obey these relations. 
Although the resulting estimator will be biased with respect to the symmetry of the bispectrum, it might conceivably have other useful properties that outweigh this consideration.
For instance, one might select estimators that are better tuned to specific features of interest in the signal bispectrum, which we will find illustrated with PAC measures.
We therefore consider next the consequence of estimating the bispectrum using different analysis filters for each band, which leads to the following estimation window:
\begin{align}
\begin{split}
\label{EQ_H}
	H(\tau,\delta) =   
	        \int{
	        		\tilde{h}_1\left(
					\frac{\nu_1+\nu_2}{2}
				 \right)\tilde{h}_2\left(\frac{\nu_2-\nu_1}{2}\right)\tilde{g}^*(\nu_2)e^{i\nu_1\delta+i\nu_2\tau}\mathop{d\nu_1}\mathop{d\nu_2}	
	     	}
\end{split}
\end{align}
with separate analysis filters, $h_1$, $h_2$, and $g$.

\begin{figure} 
\centering
\includegraphics[width=80mm]{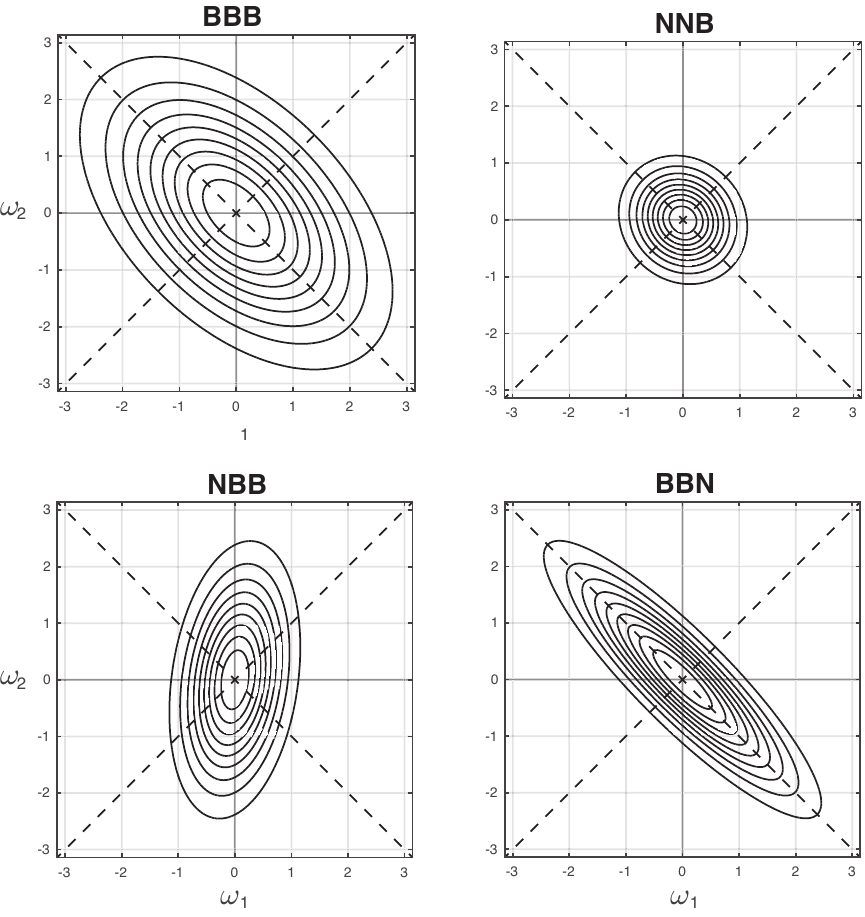}
\caption{\label{fig:kernels} 
Relationship between analysis filter bandwidth and equivalent smoothing kernels for 4 bispectral estimators.  
Dashed lines indicate the axes given by $\nu_1\equiv\omega_1-\omega_2$ and $\nu_2\equiv\omega_1+\omega_2$. In each case, one of two Gaussian analysis filters is applied to each of the three bands used in estimating the bispectrum, ordered by center frequency: $\omega_1\le\omega_2<\nu_2$. 
Frequency windows are \emph{N}arrowband with standard width, $\sigma=0.25$, or \emph{B}roadband with standard width, $\sigma=\sqrt{2}$, and the estimators are indicated according to which window was applied to each ordered band: \emph{BBB, NNB, BBN}  and \emph{BNB}.   \emph{Top Left}: \emph{BBB} (also \emph{NNN} under rescaling) Fixed analysis bandwidth. 
Each band is filtered using the same analysis window, \emph{B}. \emph{Top Right}: \emph{NNB}  Highest band is broad and the other two narrow, giving the asymptotically symmetric kernel, $\tilde{g}(\omega_1)\tilde{g}(\omega_2)$. \emph{Bottom Right}:  \emph{BBN} Highest band is narrow and the remaining two are broad giving the asymptotic kernel, $\left|h(\nu_1/2)\right|^2\tilde{g}(\nu_2)$. \emph{Bottom Left}: \emph{NBB} (also \emph{BNB} under a transpose of axes) Second band is narrow and the remaining two broad, giving the asymptotic kernel, $\tilde{h}(\omega_1)\left|g(\omega_2)\right|^2$. }
\end{figure}

In general, because the smoothing occurs over the two-dimensional bispectral plane, the choice of third window is over-constrained. This is easily recognized in the limit as the bandwidth of one filter diverges from that of the other two, in which case the smoothing window factors over the two dimensions. For example, as $\tilde{g}\rightarrow\delta$ \add{in (\ref{EQ_H}),} one obtains:
\[
	\tilde{H}(\nu_1,\nu_2) \rightarrow  
	        		\tilde{h}_1\left(
					\frac{\nu_1}{2}
				 \right)
				 \tilde{h}^*_2\left(
					\frac{\nu_1}{2}
				 \right)\tilde{g}^*(\nu_2)
\]
Whereas in the limit  $\tilde{h}_2\rightarrow\delta$,
\[
	\tilde{H}(\omega_1,\omega_2) \rightarrow  
	        		\tilde{h}_1(\omega_1)
				 \tilde{h}_2(\omega_2)\tilde{g}^*(\omega_1)
\]
and likewise for $\tilde{h}_1\rightarrow\delta$. In each of these cases the resulting window in one dimension is the product of the two broader filters. 
The estimator therefore produces the same result as when using two filters: the first given by a narrow filter and the second by the root product of the remaining two. 
Without loss of generality, the following sections therefore consider estimators for which two of the three analysis filters use the same window. 
We consider permutations of the bandwidth of the three analysis filters applied to the three bands sorted by frequency: $\omega_1\le\omega_2<\nu_2$.
Shapes of filter kernels resulting from each combination are depicted in Figure \ref{fig:kernels}.

\subsubsection{Two Filters: \textbf{N}arrow, \textbf{N}arrow, \textbf{B}road}

When $\tilde{g}$ is wider than $\tilde{h}$ \add{in (\ref{EQ_H}),} $\tilde{H}$ becomes symmetric so that $W$ is smoothed equally along the $\nu_1$ and $\nu_2$ axes. As $\tilde{g}(\nu_2) \rightarrow 1$ we obtain the approximation: 
\begin{align}
\begin{split}
\label{EQ_H1}
	\tilde{H}(\nu_1,\nu_2) \rightarrow   
	        		\tilde{h}\left(\frac{\nu_1+\nu_2}{2}\right)\tilde{h}^*\left(\frac{\nu_1-\nu_2}{2}\right)=\tilde{h}\left(\omega_1\right)\tilde{h}^*\left(\omega_2\right)
\end{split}
\end{align}
An estimate of the bispectrum with radially symmetric smoothing is therefore recovered by: 
\begin{align}
\begin{split}
\label{EQ_bispectestG}
	 B&(\omega_1,\omega_2) =  \int{\chi_h(\omega_1,t)\chi_h(\omega_2,t)X(t)e^{i(\omega_1+\omega_2)t}\mathop{dt}
			}\\
	\end{split}
\end{align}
Note that the radial symmetry here does not obey the symmetry relations of (\ref{EQ_symmetry}), so the estimate does not exactly recover the symmetry of the bispectrum. 
On the other hand, (\ref{EQ_bispectestG}) suggests a less cumbersome approach to estimation which foregoes the summation over the product of three separate bands. 
The estimator is obtained instead from the frequency-domain covariance of the time-frequency decomposition after remodulation and weighting by the original signal. 
In practice, however, the time-frequency decomposition is often downsampled, so it will still be necessary to either apply an anti-aliasing filter to $X$ appropriate for the sampling rate and frequency range under consideration, or oversample $\chi_h$. Both options tend to negate any computational advantage that might otherwise be gained from such a simplification.

\subsubsection{Two filters: \textbf{B}road, \textbf{B}road, \textbf{N}arrow}

The estimator in Eq. (\ref{EQ_bispectestG}) results in a smoothing window with radial symmetry, but the symmetry came about from greater smoothing in the $\nu_2$ direction entailing a sacrifice of resolution.  One may consider the converse possibility, allowing the bandwidth of $g$ \add{(in \ref{EQ_H})} to be narrower than that of $h$. 
This produces a smoothing window with the limit $\tilde{g}(\nu_2) \rightarrow \delta(\nu_2)$. In the approach to this limit one obtains the following approximation:
\begin{align}
\begin{split}
\label{EQ_HG2}
	\tilde{H}(\nu_1,\nu_2) \rightarrow   
	        		\left|\tilde{h}\left(
					\frac{\nu_1}{2}
				 \right)\right|^2\tilde{g}^*(\nu_2)
\end{split}
\end{align}
Provided that $g$ is narrower than $h$, this estimator allows the smoothing along $\nu_1$ and $\nu_2$ to be separately controlled.

\subsubsection{Two filters: \textbf{N}arrow, \textbf{B}road, \textbf{B}road}
\label{sec:nbb}
In the limit $\tilde{h}_1 \rightarrow \delta$ \add{in (\ref{EQ_H}),} we have $g(\nu_2)\rightarrow g(\omega_2)$. Letting $h_2=g$, without loss of generality, we have
\begin{align}
\begin{split}
\label{EQ_Hnbb}
	\tilde{H}(\omega_1,\omega_2) \rightarrow   
	        		\tilde{h}\left(
					\omega_1
				 \right)\left|\tilde{g}(\omega_2)\right|^2
\end{split}
\end{align}
The same result holds for the BNB estimator swapping the arguments. The smoothing kernel here is related to the previous examples through a 45 degree rotation in the bispectral plane. One important potential drawback of this family of estimators is that the outcome will not reflect the symmetry of the bispectrum about the diagonal, $\omega_1=\omega_2$. Measures of phase-amplitude coupling relate to bispectral estimators of this type, as shown in \ref{appendix:PhPC_proof}.

\section{Phase-Power Coherence as a Smoothed Bispectral Estimator}
\label{appendix:PhPC_proof}

The following appendix provides \addTwo{a} mathematical demonstration of the equivalence between phase-power coherence (PhPC) and the NBB class of bispectral estimators. 
PhPC is calculated from the cross spectrum between the squared analytic envelope at one frequency and the original signal at another.
\addTwo{The goal is normally to isolate the fast oscillation within the first band and the slow oscillation within the second.} 
The squared  envelope \delTwo{a band}\addTwo{of the FO band,} centered at $\gamma$ filtered with the analysis window, $h$, is given by
\begin{align}
\label{EQ_powerenv}
	P(\gamma,t) =  \iint{
				h(r-t)h^*(u-t)X(r)X^*(u)e^{-i\gamma\left(r-u\right)}
				\mathop{dr}\mathop{du} 
			}
\end{align}
PhPC is calculated from the cross spectrum between $P$ and $X$, obtained with a second analysis filter, $g$, which will be of a narrower band than $h$\addTwo{, targeted to the range of the SO} around frequency $\theta$:
\begin{align}
\label{EQ_pac1}
\begin{split}
	\phi(\theta,\gamma) &=  \iiint{
				g(t-\tau)g^*(s-\tau)X(s)P^*(\gamma,t)e^{
							-i\theta(s-t)
					}
								\mathop{ds} \mathop{dt} \mathop{d\tau} 
			}\\
			&=  \iint{
				g^{(2)}(s-t)X(s)P^*(\gamma,t)e^{-i\theta(s-t)}
								\mathop{ds} \mathop{dt} 
			}
\end{split}
\end{align}
where $g^{(2)}$ denotes the autocorrelation of $g$. Expanding $P$ \add{in (\ref{EQ_pac1}) according to (\ref{EQ_powerenv}),} we have
\begin{align}
\label{EQ_pac}
\begin{split}
	&\phi(\theta,\gamma) =\\
			 &  \iiiint{
				g^{(2)}(s-t)h(r-t)h(u-t)X(s)X(r)X^*(u)e^{
						-i\gamma(r-u) - i\theta(s-t)
						}
								\mathop{ds} \mathop{dt} \mathop{du} \mathop{dr} 
				}\\
\end{split}
\end{align}
\add{Following the foregoing series of substitutions, the integral over $t$ \add{in (\ref{EQ_pac})} no longer depends on the signal (which is now indexed by $r$,$s$ and $u$). Pulling out} the term that integrates over $t$ \del{independently of the signal }yields a smoothing kernel:
\begin{align}
\label{EQ_W}
\begin{split}
	A
			=&  \int{
				g^{(2)}(s-t)h(r-t)h(u-t)e^{-i\theta(s-t)}\mathop{dt}  
				}\\
			=&  \int{
				\left|\tilde{g}(\xi_3)\right|^2\tilde{h}(\xi_1)\tilde{h}^*(\xi_2)e^{i\xi_1r - i\xi_2u + i(\xi_3-\theta)s -i(\xi_1 - \xi_2+\xi_3-\theta )t}\mathop{d\xi_1}\mathop{d\xi_2}\mathop{d\xi_3}\mathop{dt}  
				}\\
			=&  \int{
				\left|\tilde{g}(\xi_2-\xi_1+\theta)\right|^2\tilde{h}(\xi_1)\tilde{h}^*(\xi_2)e^{i\xi_1(r -s)-i\xi_2(u -s)}\mathop{d\xi_1}\mathop{d\xi_2}				}\\
			=&  \int{
				\left|\tilde{g}(\theta+2\lambda_2)\right|^2\tilde{h}\left(\lambda_1+\lambda_2\right)\tilde{h}^*\left(\lambda_1-\lambda_2\right)e^{i\lambda_1(r-u) - i\lambda_2(r+u) + i2\lambda_2s }\mathop{d\lambda_1}\mathop{d\lambda_2} 
				}\\
\end{split}
\end{align}
from which, setting $r = s+\tau_2$ and $u = s + \tau_1$:
\begin{align}
\label{EQ_W2}
\begin{split}
	A
		      =&  \int{
				\left|\tilde{g}(\theta+2\lambda_2)\right|^2\tilde{h}\left(\lambda_1+\lambda_2\right)\tilde{h}^*\left(\lambda_1-\lambda_2\right)e^{-i(\lambda_1+\lambda_2)\tau_1 - i(\lambda_2-\lambda_1)\tau_2 }\mathop{d\lambda_1}\mathop{d\lambda_2} 
				}\\
\end{split}
\end{align}
%
Substituting back into (\ref{EQ_pac}) \add{and applying (\ref{EQ_ambi3rd})}, one arrives at 
\begin{align}
\label{EQ_pac3}
\begin{split}
	\phi(\theta,\gamma) &=\\
	& \iiint{
				\left|\tilde{g}(\theta+2\lambda_2)\right|^2\tilde{h}\left(\lambda_1+\lambda_2\right)\tilde{h}^*\left(\lambda_1-\lambda_2\right)W(\lambda_2+\lambda_1-\gamma,\lambda_2-\lambda_1+\gamma,z)
								\mathop{d\lambda_1}\mathop{d\lambda_2}\mathop{dz} 
				}\\
	\end{split}
\end{align}
which with another change of variables, $\omega_1\equiv -2\lambda_2$ and $\omega_2\equiv\lambda_2-\lambda_1+\gamma$ and applying the symmetry relation from Eq. (\ref{EQ_Fsymmetry}),  $W(-\omega_2-\omega_1,\omega_2,z)=W(\omega_1,\omega_2,z)$, this becomes
\begin{align}
\label{EQ_pac4}
\begin{split}
	\phi(\theta,\gamma) &= \\ 
	  &\iiint{
				\left|\tilde{g}\left(\omega_1-\theta\right)\right|^2\tilde{h}^*\left(\omega_2+\omega_1-\gamma\right)\tilde{h}\left( \omega_2-\gamma \right)W(\omega_1,\omega_2,z)
								\mathop{d\omega_1}\mathop{d\omega_2}\mathop{dz} 
			}\\
	\end{split}
\end{align}
Finally, because in measuring phase-amplitude coupling, the SO filter, $g$, is selected with a narrower bandwidth than the FO filter, $h$, approaching the limit $\tilde{g}\rightarrow\delta$, we may use the following good approximation
\begin{align}
\label{EQ_pac5}
\begin{split}
	\phi(\theta,\gamma) &\approx \\ 
	  &\iiint{
				\left|\tilde{g}\left(\omega_1-\theta\right)\right|^2\tilde{h}^*\left(\omega_2+\theta-\gamma\right)\tilde{h}\left(\omega_2-\gamma\right)W(\omega_1,\omega_2,z)
								\mathop{d\omega_1}\mathop{d\omega_2}\mathop{dz} 
				}\\
\end{split}
\end{align}

\subsection{Relationship between PhPC  and the Bispectrum}
\label{section:PhPC_bispectrum}
\add{The equivalent smoothing window for PhPC estimators, given by} Eq. (\ref{EQ_pac5}), very nearly describes an NBB smoothing kernel like the one in (\ref{EQ_Hnbb}). 
\add{The only nontrivial difference arises from the fact that the smoothing kernel varies as a function of the SO frequency,} $\theta$\add{.}\del{as an argument in one of the broad window, $h$, terms.} The effect \del{of this term is easily seen} is easy to appreciate for the Gaussian window of bandwidth $\sigma_\omega$:
 \begin{align}
	\tilde{h}^*\left(\omega_2-\gamma+\theta\right)\tilde{h}\left(\omega_2-\gamma\right)=
	e^{-\frac{1}{\sigma^2_\omega}\left[2\left(\omega_2-\gamma+\theta/2\right)^2 - \theta^2/2\right]}
\end{align}
For the Gaussian case, the result is the same as smoothing with an NBB window, only shifted along $\omega_2$ by $\theta/2$ and attenuated as a function of the phase-providing frequency by $e^{-\theta^2/2}$ without any other change in the size or shape of the smoothing window.  More generally, if $h$ is symmetric, then $\tilde{h}^*\left(\omega_2-\gamma+\theta\right)\tilde{h}\left(\omega_2-\gamma\right)$ is symmetric and centered at $\gamma - \theta/2$. 

PAC measures require the FO filter to have a bandwidth at least as great as the range of SO frequencies of interest, because the spectral range of fluctuations in the power amplitude is limited to the bandwidth of the filter used to extract them, so that $\theta$ must remain small relative to the bandwidth of $h$ and $\left|h(\omega+\theta)\right|\approx \left|h(\omega)\right|$\cite{aru2015untangling}. 
Precisely within this range, PAC estimates may be treated directly as NBB estimators of the bispectrum, which use the window:  
\begin{align}
\label{EQ_HpacApprox}
	H_{PAC} =\left|\tilde{g}\left(\omega_1\right)\tilde{h}\left(\omega_2\right)\right|^2	
\end{align}
This constraint that limits the spectral range of the SO according to the bandwidth of the FO seems at first glance to be a fundamental consideration in measuring PAC, but the preceding analysis shows it to be  in fact a dispensable property of the estimator unrelated to anything inherent in the underlying measured quantity.

One approach to circumventing the spectral limitation of PAC estimators modifies the FO filters used according to the SO frequency. For example, the problem is sometimes addressed by scaling the bandwidth of the FO filter, $h$, by the center frequency of the SO filter \citep{berman2012variable}:
$\tilde{h}_\theta(\omega)=\tilde{h}\left(\frac{\omega}{\theta}\right)$, giving
\begin{align}
\label{EQ_HpacScale}
\begin{split}
\tilde{h}_\theta^*\left(\omega_2+\theta-\gamma\right)\tilde{h}_\theta\left(\omega_2-\gamma\right) 
=& \tilde{h}^*\left(\frac{\omega_2-\gamma}{\theta}+\frac{1}{2}\right)\tilde{h}\left(\frac{\omega_2-\gamma}{\theta}\right)\\
\approx & \left|\tilde{h}\left(\frac{\omega_2-\gamma}{\theta}\right)\right|^2
\end{split}
\end{align}
This adjustment removes the attenuation with SO frequency by ensuring that the bandwidth of the FO filter is appropriate for modulation frequencies in the range of $\theta$.
It still results in diminishing resolution of the FO with increasing SO, which is not a necessary constraint on bispectral estimators.

An apparent disadvantage of standard bispectral estimators might be noted at this point, which is the shift of the FO spectrum by $\theta/2$.
This shift from the FO frequency is however not essential to the bispectral estimator and can be removed through the symmetrizing change of variables given in Eq. (\ref{EQ_bispectsym}).

It should be reemphasized that these differences between PhPC and bispectral estimators involve rather arbitrary properties of the respective smoothing kernels and do not reflect any essential difference in the quantity measured.  
It might be argued that the limit on the combined range and resolution of the estimator imposed by PAC helps tune it to bispectral characteristics of PAC. 
Such tuning may improve signal-to-noise sensitivity to PAC as well as computational efficiency, but it nevertheless forces some \emph{a priori} assumptions about the bandwidth and spectral range of PAC.
More importantly, smoothing may sacrifice information vital for distinguishing between alternative forms of cross-frequency coupling.

\subsection{Interregional PAC and Cross Bicoherence}

The foregoing discussion extends naturally to bivariate cross-frequency coupling involving two signals, for example, obtained from two separate recording channels.
The bivariate extension of PAC measures describe\addTwo{s} the dependence between amplitude in one signal with phase in another. Using the same argument as in the single-channel case, measures of interregional PAC may be equated to the cross bispectrum of the form:
\begin{align}
\label{EQ_crossbicoh}
	 B_{ij}(\omega_1,\omega_2) =  \mathrm{E}\left[ \tilde{X}_i(\omega_1)\tilde{X}_j(\omega_2-\frac{\omega_1}{2})\tilde{X}_j^*(\omega_2 + \frac{\omega_1}{2})\right]
\end{align}
where in the language of PAC, $i$ indexes the ``phase-providing'' signal and $j$ the ``amplitude-providing'' signal.


 \section{Transient Signal Model}
 \label{section:signal_model}
The following appendix presents a basic signal model that will serve as a starting point in describing bispectral features associated with phase-amplitude coupling and other third-order signal features. 
It describes the case when recurring transient features with characteristic but unknown spectra lie embedded at unknown times in the observed record. 
The generating process can be separated into two parts: a (1) \emph{point process}, $N$, whose increments determine the times, $\tau_i$, at which (2) some random or deterministic transient feature, $f_i(t)$ is embedded in the signal. 
The first process is concerned only with timing and the second only with the emitted waveform.
The spectral representation of such a signal is
 \begin{align}
\label{EQ_PPF0}
\begin{split}
			\tilde{X}(\omega)&=\sum_i{\tilde{f}_i(\omega)e^{-i\omega\tau_i}}
\end{split}
\end{align}
The following sections consider the bispectrum that results from this process when $f_i$ is independent of time and of N and, in the first case, deterministic, and, in the second, i.i.d. random.
\subsection{Deterministic Features}
If the $f_i$ \add{in (\ref{EQ_PPF0})} are deterministic with $f_i=f$, it follows from the convolution property (Eq. \ref{EQ_convprop})
\begin{align}
\label{EQ_PPBC}
\begin{split}
			 B_X(\omega_1,\omega_2)&=\tilde{f}(\omega_1)\tilde{f}(\omega_2)\tilde{f}^*(\omega_1+\omega_2)B_N(\omega_1,\omega_2)\\
\end{split}
\end{align}
with
\begin{align}
\label{EQ_PPBN}
\begin{split}
			 B_N(\omega_1,\omega_2) = \iiint{e^{-i\omega_1r-i\omega_2s+i(\omega_1+\omega_2)t }\mathrm{E}\left[\mathop{dN(r)}\mathop{dN(s)}\mathop{dN(t)}\right]
			}
\end{split}
\end{align}
For a ``simple'' point process, events coincide with vanishing probability, so that $E[(dN(t))^k] = E[dN(t)]$. For a process with the first three moments defined, $\mu_1$, $\mu_2$, $\mu_3$ \citep{bartlett1963spectral,daley2003introductionI}, the expectation in Eq. (\ref{EQ_PPBN})  yields
\begin{align}
\label{EQ_Poiss}
\begin{split}
	\mathrm{E}\left[\mathop{dN(r)}\mathop{dN(s)}\mathop{dN(t)}\right] =&\\
	 &\left[\mu_1(t)\delta(r-t)\delta(s-t)\right.\\
	 &+\mu_2(r-t)\delta(s-t)+\mu_2(s-t)\delta(r-t)+\mu_2(s-t)\delta(r-s)\\
	 &\left.+\mu_3(r-t,s-t)\right]\mathop{dr}\mathop{ds}\mathop{dt}
\end{split}
\end{align}
The bispectrum for this process is then
\begin{align}
\label{EQ_FPoiss}
\begin{split}
	B_N(\omega_1,\omega_2)=\lambda\left[ 1 + \tilde{\mu}_2(\omega_1)+\tilde{\mu}_2(\omega_2) +\tilde{\mu}_2(\omega_1+\omega_2) + \tilde{\mu}_3(\omega_1,\omega_2)\right]
\end{split}
\end{align}
where $\lambda$ is the average rate ($N/T \rightarrow  \lambda$).

For a stationary homogenous Poisson driving process (having constant $\mu(t) = \lambda$), \add{Eq. (\ref{EQ_FPoiss})} reduces to 
\begin{align}
B_N(\omega_1,\omega_2)=\lambda + \lambda^2\left[\delta(\omega_1)+\delta(\omega_2)+\delta(\omega_1+\omega_2)\right] + \lambda^3\delta(\omega_1)\delta(\omega_2)  
\end{align} 
so that, neglecting the probability mass at the origins and along $\omega_1+\omega_2=0$,
\begin{align}
B_X=\lambda \tilde{f}(\omega_1) \tilde{f}(\omega_2) \tilde{f}^*(\omega_1+\omega_2)
\end{align}
This quantity is nonzero when $f$ contains harmonics or is otherwise spectrally broad, so that its support covers some $\omega_1,\omega_2$, and $\omega_1+\omega_2$.
In fact, because $f$ is  transient by assumption, this condition is already given: $f$ must decay to zero within some finite time window, $\Delta T$, and therefore its spectrum contains at a minimum a main lobe and side lobes spaced at $2\pi/\Delta T$. But if $f$ has zero mean, a high center frequency and is highly oscillatory with an otherwise narrow spectrum, the product may still be negligible. The bispectrum will contain larger peaks when $f$ has a broad spectrum on the order of 2/3 its center frequency,  non-zero mean, or contains one or more harmonic complexes.

\subsection{Random Features}
\label{section:random_features}
Extending the preceding analysis to the case when each $f_i$ \add{in (\ref{EQ_PPF0})} is itself the realization of a random process independent of $N$,
\begin{align}
\label{EQ_PPBstochastic}
\begin{split}
			 B_X(\omega_1,\omega_2) = &\sum_{ijk}{
			  \mathrm{E}\left[\tilde{f}_i(\omega_1)\tilde{f}_j(\omega_2)\tilde{f}^*_k(\omega_1+\omega_2)e^{i\omega_1(\tau_k-\tau_i)+i\omega_2(\tau_k-\tau_j) }\right]
			 }\\
			  = & \mathrm{E}\left[
			 \sum_i{\tilde{f}_i(\omega_1)\tilde{f}_i(\omega_2)\tilde{f}^*_i(\omega_1+\omega_2)}\right.\\
			 &+ \sum_{i\ne j}{ \tilde{f}_i(\omega_1) \tilde{f}_j(\omega_2)\tilde{f}^*_j(\omega_1+\omega_2)e^{-i\omega_1(\tau_i-\tau_j)}}\\
			  &+ \sum_{i\ne j}{\tilde{f}_i(\omega_2)\tilde{f}_j(\omega_1)\tilde{f}^*_j(\omega_1+\omega_2)e^{-i\omega_2(\tau_i-\tau_j)}  }\\
			 	 &+ \sum_{i\ne j}{\tilde{f}_i(\omega_1)\tilde{f}_i(\omega_2) \tilde{f}^*_j(\omega_1+\omega_2) e^{-i(\omega_1+\omega_2)(\tau_i-\tau_j)}}\\
			&+\left. \sum_{i\ne j\ne k}{ \tilde{f}_i(\omega_1) \tilde{f}_j(\omega_2) \tilde{f}^*_k(\omega_1+\omega_2) e^{-i\omega_1(\tau_i-\tau_k)-\omega_2(\tau_j-\tau_k)}}\right]\\
\end{split}
\end{align}
 If the process generating $f_i$ is independent of $N$ and time, so that $\tilde{f}_i\sim\tilde{f}_j$, for all $i,j$, and $N$ simple, with the first three moments defined, $\mu_1$, $\mu_2$, $\mu_3$ \citep{bartlett1963spectral,daley2003introductionI}:
\begin{align}
\label{EQ_PPBC2}
\begin{split}
			 B_X(\omega_1,\omega_2)=
			  &
			\lambda\left[\left<\tilde{f}(\omega_1)\tilde{f}(\omega_2)\tilde{f}^*(\omega_1+\omega_2)\right>\right.\\
			 &+ \left<\tilde{f}(\omega_1)\right>
			 \left<\tilde{f}(\omega_2)\tilde{f}^*(\omega_1+\omega_2)\right>
			  \tilde{\mu}_2(\omega_1)\\
			  &+ \left<\tilde{f}(\omega_2)\right>
			 \left<\tilde{f}(\omega_1)\tilde{f}^*(\omega_1+\omega_2)\right>
			  \tilde{\mu}_2(\omega_2)\\
			 &+ \left<\tilde{f}(\omega_1)\tilde{f}(\omega_2)\right>
			 \left<\tilde{f}^*(\omega_1+\omega_2)\right>
			 \tilde{\mu}_2(\omega_1+\omega_2)\\
			 &+ \left.\left<\tilde{f}(\omega_1)\right>\left<\tilde{f}(\omega_2)\right>\left<\tilde{f}^*(\omega_1+\omega_2)\right> \tilde{\mu}_3(\omega_1,\omega_2)\right]\\
\end{split}
\end{align}
with the brackets denoting expectation, and a normalization by the cumulative event count, $N$, is left implicit. This reduces to $B_X=B_fB_N$ if the error of $f_i$ involves independent additive Gaussian noise.  

This division of labor allows one to separate the spectral contributions of local features described by $f$ from those of the large-scale driving process; these respective contributions can be distinguished in the bispectrum\footnotemark. In the case of phase-amplitude coupling, $f$ will be treated as the sum of two components: a single burst of fast oscillations, $f_{\mathrm{FO}}$, whose spectrum is characteristically broad and smooth, and a slow oscillation, $f_{\mathrm{SO}}$, with a narrow spectrum. In constructing the full signal, narrowband large-scale features can be viewed as the consequence of a filter applied to the  train of impulses generated by the point process, where $f_{\mathrm{SO}}$ plays the role of filter function. 

\footnotetext{The Poisson formulation adopted here for the driving point process is mildly restrictive in that it assumes the event count within a given interval follows a Poisson distribution.  Periodicity in the driving process therefore does not entail a train of evenly spaced events, but instead a train of evenly spaced event clusters with Poisson-distributed size.  An alternative point-process model that conditions on history, such as a renewal process, might handle periodicity more naturally in this setting, but elucidating the spectra of such processes is more technically involved, making them less useful for the present purpose.  One way to address the case of simple uniform periodicity with the Poisson model, which leaves the foregoing spectral analysis unchanged, is to allow $\lambda$ to take the form of transient bursts of infinitesimal duration, each burst generating a cluster with Poisson-distributed size. The resulting signal then contains $f$ at the cluster times modulated by cluster size.  The variance of amplitude in $f$ introduced by this weighting with cluster sizes can be made arbitrarily small by scaling $\lambda$ so as to increase cluster size, with the inverse scaling applied $f$ to keep amplitude fixed. }

\end{appendix}
\end{document}